\documentstyle[aps,pre,epsfig]{revtex}

\begin{document} 

\draft
 
\title{Mechanisms for slow strengthening in granular materials} 
 
\author{W. Losert$^{1}$, J.-C. G\'eminard$^{1,2}$, S. Nasuno$^{1,3}$, 
and J.P. Gollub$^{1,4}$} 

\address{$^{1}$ Physics Department, Haverford College, Haverford, PA 19041}

\address{$^{2}$ Laboratoire de Physique de l'E.N.S. de Lyon, 46 All\'ee 
d'Italie, 
69364 Lyon Cedex, France}
\address{$^{3}$ Department of Electrical Engineering, 
Kyushu Institute of Technology,
Tobata, Kitakyushu 804-8550, Japan}
\address{$^{4}$ Physics Department, University of Pennsylvania,
Philadelphia, PA 19104}

\date{\today}

\maketitle

\begin{abstract}
Several mechanisms cause a granular 
material to strengthen over time at low applied stress.
The strength is determined from the maximum frictional force 
$F_{max}$ experienced by a shearing plate in contact with 
wet or dry granular material after 
the layer has been at rest for a waiting time $\tau$. 
The layer strength increases roughly logarithmically 
with $\tau$ {\it only} if a shear stress is applied 
during the waiting time. 
The mechanisms of strengthening
are investigated by sensitive displacement measurements and by imaging
of particle motion in the shear zone.
Granular matter can strengthen due to a 
slow shift in the particle arrangement under 
shear stress.
Humidity also leads to strengthening, 
but is found not to be its sole cause. 
In addition to these time dependent effects,
the static friction coefficient
can also be increased by compaction of the granular material 
under some circumstances, and by cycling of the applied shear stress.
\end{abstract}
 

\section{INTRODUCTION} 
The strength of granular matter is an important macroscopic 
property.  Under many circumstances a layer of granular material at rest
can sustain a load, i.e. it behaves like a solid.  However, applied shear
forces can lead to partial fluidization, a phenomenon that
 has no direct analog in homogeneous materials~\cite{Nagel96}. 
The threshold shear stress $\sigma_{Max}$ needed to initiate flow 
also determines 
when a granular material will lose much of its ability to sustain 
loads, and is therefore a good measure of strength.

The shear strength of a granular layer is of great interest in geophysical
applications, due to the granular composition of earthquake faults.
Various studies (described in section \ref{background})
have shown a gradual strengthening of a granular layer
at geophysical pressures, as the time between stress relieving events
is increased. Since strengthening of geophysical faults
 can affect the temporal distribution of earthquakes (see Ref.~\cite{Marone98a}
for references),
understanding the underlying mechanisms and developing appropriate
models is of practical interest~\cite{Scholz98}.

In previous studies by our  group~\cite{Gollub98,Nasuno97,Geminard99} 
the frictional properties of wet and dry sheared granular materials 
at low pressures
were determined in detail by shearing the layer by means of a plate resting 
upon it. At small applied 
stresses,  high resolution stress and displacement measurements
can provide an accurate determination of the instantaneous
frictional force experienced by the plate.
The static friction coefficient $\mu_s$ (the ratio of the threshold static
shear stress $\sigma_{Max}$ to normal stress) 
was found to be reproducible under given experimental conditions. 
However, changes in the experimental history influence $\mu_s$.
In this article we study the effect of the experimental history on 
the threshold static shear stress 
of granular material at low applied normal stress. 
One central parameter of the experimental history is the 
waiting time $\tau$ for which the layer has been at rest. 
In addition, the shear stress and humidity during 
the waiting time are varied; and measurements are made in fluid saturated 
material. The effects of applied stress reversal are also studied.

In order to determine mechanisms for strengthening, we measure
 the instananeous frictional force and the instantaneous vertical dilation of 
the layer during waiting and during motion.
We also image a cross section of the granular layer in some experiments
and track the motion of individual grains during waiting and during 
motion of the plate.  Our results indicate that several different
mechanisms are needed 
to account for the observed strengthening.  
We find that granular matter can strengthen due to a 
slow shift in the particle arrangement under 
shear stress. This observation is consistent with a model for granular
matter as consisting of a fragile network of stress chains that adjusts
to the applied stress~\cite{Cates98}.
Strengthening is also influenced by humidity, 
since condensation of liquid bridges
adds an attractive surface tension force between grains~\cite{Ciliberto98}. 
We show clearly that this cannot be the only
cause of strengthening by carrying out experiments under water, 
where liquid bridges cannot be formed.
Finally, strengthening may occur by the evolution of individual 
microcontacts through slow creep~\cite{Baumberger99a,Baumberger99b}. 
Our experiments are not sufficiently sensitive to determine whether 
creep on length scales of the size of microcontacts actually occurs.  
However, strengthening of individual 
contacts alone cannot account for the experimental results.
In addition to these slow time-dependent effects,
the static friction coefficient
can also be increased by compaction of the granular material 
under some circumstances, and by cycling of the applied shear stress.

Relevant work on friction and work related to proposed
 strengthening mechanisms are discussed in 
section~\ref{background}.
The experimental results are described in 
sections~\ref{expdata} and
discussed in section~\ref{discussion}.

\section{BACKGROUND} \label{background}

\subsection{Strengthening in solid-on-solid friction}

In friction between dry solid surfaces (solid-on-solid friction) the static 
friction coefficient $\mu_s$ is determined
by the number and strength of microcontacts that support the applied 
stress (for recent reviews on friction, see Refs.~\cite{Haehner98,Krim98}).
Since the strength of microcontacts changes with time or applied stress,
and since their number changes as well through stress-induced creep, 
$\mu_s$ is not a constant, but is influenced by the state of the 
system.
The static friction coefficient  - the ratio of the shear force needed to 
initiate sliding to the normal force - 
has generally been found to increase approximately 
logarithmically with the time of (quasi)-stationary contact between 
materials. 

The static strength of a glass-on-glass interface was recently
investigated experimentally by Berthoud 
{\it et al.}~\cite{Baumberger99a,Baumberger99b}.
In addition to logarithmic strengthening, the authors find that the rate
of increase grows with the temperature of the material.
The strengthening rate is twice as high when
a shear stress is applied during the waiting time, but strengthening 
persists without an applied shear stress.
This strengthening was found to be a result of an increase in the load bearing 
area through load-induced creep of individual microcontacts.

\subsection{Strengthening in geophysical experiments}

In studies at geophysical pressures ($\sim 20 {\rm MPa}$)
using simulated fault gouge (granular quartz powder)
Marone {\it et al.}~\cite{Marone98b,Marone98a} found logarithmic strengthening
of a granular layer with time, if shear stress is applied during the 
waiting time. On the other hand, when the shear stress is removed,
an immediate strengthening of the material is observed followed by
a slow weakening with waiting time.
Nakatani {\it et al.}~\cite{Nakatani98} showed that the 
magnitude of the instantaneous strengthening is proportional to the
amount by which the applied shear stress is reduced.
These effects are suggested to be a consequence of 
'overconsolidation' i.e. rearrangement of particles into a more
compact configuration.

\subsection{Rate and state theories}
In order to describe the dependence of the frictional force on the 
state of the system, rate and state theories were developed
by Dieterich~\cite{Dieterich79} and Ruina~\cite{Ruina83}, 
which introduce an additional variable $\Theta$ that characterises the system
state. In general, several variables might be necessary to
describe the state of the system, but the two models focus on 
the simplest possible case of one state variable determining the
system state.
In both theories the frictional force is given by
\begin{equation}
\mu = \mu_0+a \ln\frac{V}{V_0} + b \ln \frac{V_0 \Theta}{D_c}
\end{equation}
where $\mu_0$, $V_0$, and $D_c$ are  characteristic constants of the
materials, and $V$ is the instantaneous sliding velocity.
In this approach one does not distinguish between static and dynamic 
cases. The models differ in the differential equation for
 the state variable $\Theta$.
In the Ruina model, for which 
$d\Theta / dt =- V \Theta / D_c \ln \left[V / \Theta D_c \right] $, 
creep is necessary ($V \ne 0$) to change
$\Theta$. On the other hand, time of contact alone increases $\Theta$
and thereby $\mu$ in the Dieterich model, where 
$d\Theta/dt = 1- V / \Theta D_c$.
Different equations have
also been proposed by Nielsen, Carlson and Olsen~\cite{Nielsen99}
to describe earthquake faults. In their model, $\Theta$ changes
both with the time of contact and with creep
using a characteristic strengthening time $\tau_c$ and 
a characteristic weakening length of contacts $l_c$
($d\Theta/dt = (1-\Theta) / \tau_c - \Theta V / l_c$).
The friction coefficient is determined by $\Theta$ and increases 
with velocity with a viscous coefficient $\eta$ ($\mu = \Theta + \eta V$).

While they successfully explain important earthquake 
characteristics~\cite{Scholz98},
these rate and state models do not describe the instantaneous
strengthening when the shear stress is released during a waiting
time~\cite{Marone98a,Nakatani98}.

\subsection{Force transmission in granular media}

In solid-on-solid friction, the microcontacts 
form a two-dimensional disordered array at the interface between the solids.
In granular materials, however, simulations indicate that 
several layers of grains
move during the shearing 
process~\cite{Hanes85,Thompson91,Aharonov98,Morgan99a}. 
This is not entirely surprising, given that 
the interior of the material is not stronger than the surface region, 
as long as there are no cohesive forces between particles.
In order to study granular friction theoretically, one therefore must
analyse a three dimensional network of microcontacts. 
The failure of a subset of
those contacts allows motion of the shearing plate.
Distinct element method simulations by Morgan 
{\it et al.}~\cite{Morgan99a,Morgan99b} indicate
that the grain size distribution and interparticle frictional properties 
can strongly affect the shear stress necessary for a failure of that
contact network and thus granular flow.

Microcontacts can be characterized by the total force and shear force
on the contact, and the orientation of the microcontact 
relative to the total applied stress.
Recent experiments have shown that the distribution and orientation of
individual particle contacts is highly nonuniform and hysteretic.
Shear and normal stresses are transmitted along stress chains, as 
for example demonstrated in recent simulations by 
Radjai {\it et al}~\cite{Radjai98} and in experiments using
birefringent disks by Howell {\it et al.}~\cite{Howell99}.
The magnitude of stress between particles follows an exponential
probability distribution~\cite{Howell99} unless the packing fraction 
exceeds a critical threshold.
The arrangements of stress chains strongly depends on the history of the
sample. Small particle rearrangements, e.g. through heating of
an individual particle, can alter the stress transmission~\cite{Nagel94}.
Recent experiments by Vanel {\it et al.}~\cite{Behringer99} have revealed that
the stresses at the bottom of a sandpile can have either a maximum
or a local minimum at the center,
depending on the preparation procedure. 
This experiment indicates that even the preferential direction of 
stress chains may be affected by the history of the sample.
One model of stress transmission by Cates {\it et al.}~\cite{Cates98}
does not try to develop a stress-strain relations,
but instead relates different components of the stress tensor
to each other.
The granular material is described as a fragile network of stress
transmitting contacts. If the direction of the applied stress changes,
the network breaks.
Simulations by Radjai {\it et al.}~\cite{Radjai98} 
suggest that the network of stress chains
can be separated into a strong subnetwork in which most
contact surfaces are perpendicular to the direction of 
total stress, and a weak subnetwork of contacts oriented preferentally
perpendicular to it.  The weak subnetwork sustains shear stress
and breaks through frictional sliding between bead surfaces.
The strong subnetwork sustains the normal load, and its contacts change mostly
through rolling. 

\subsection{Role of dilation}

It has been known since Reynolds~\cite{Reynolds} 
that shearing a granular material
produces dilation, i.e. an expansion of the material in
the direction perpendicular to the motion. 
In wet granular materials, our group found that a granular
layer dilates by $10\%$ of one grain diameter
while the top layer is translated by one grain 
diameter~\cite{Geminard99}; this observation suggests 
that at most a few layers are involved in the dilation.  
Computer simulations by Thompson and Grest~\cite{Thompson91} and
Zhang and Campbell~\cite{Campbell92} have indicated 
that dilation is prominent in stick-slip motion. 
A region of about 6-12 grain diameters typically dilates and starts to flow
in the simulations at small normal forces. 
Zhang and Campbell found an approximately linear decrease in particle 
velocity with depth at high normal forces, and a faster than linear decrease
at low normal forces.

Experiments at geophysical pressures 
by Marone have shown that the granular layer  
becomes compacted during the waiting time at high normal 
stresses~\cite{Marone98a}.   
Marone notes that compaction increases in proportion to the frictional
strength of the material; this indicates a close relation between compaction
and strength. The role of dilation therefore should be investigated
in connection with measurements of the frictional strength.

\subsection{Role of humidity}

Experiments in a rotating cylinder~\cite{Ciliberto98} have indicated that 
strengthening of granular material (indicated by an increased critical
angle of repose)
might be due to the formation of liquid bridges,
which introduce attractive capillary forces between grains.
A logarithmic increase of the critical
angle of repose of the pile with waiting time
was observed. The rate of increase was found to depend strongly on humidity 
and no strengthening occurred for bead diameters $d>500 \mu m$ or
for vanishing humidity.
This behavior is consistent
 with liquid bridges being the sole cause of strengthening in
this study, but angle of repose
experiments only look at strengthening in the limit of very small stress, 
where the role of other strengthening mechanisms might be diminished.

We varied the humidity in some experiments and 
carried out experiments under water, where liquid bridges cannot form,
in order to distinguish this mechanism 
of strengthening from other possibilities.

\subsection{This work on strengthening in granular materials}

Judging from the three dimensional nature of the sheared region, the
role of dilation, humidity, and the nonuniform and hysteretic transmission
of stress through a granular material, 
we cannot expect {\it a priori} that the laws that govern
strengthening of solids will hold for granular materials.

In our experiments a small, but non-negligible normal force is
applied to the granular material, in contrast to the strengthening
studies based on the critical angle of repose. On
the other hand, the applied normal force is smaller than the one used
in geophysical experiments by a factor of about $10^6$.  
The threshold stress for plastic deformation of the glass particles is reached
for an approximate area for individual microcontacts 
of the order of $25~{\rm nm^2}$ per particle in our 
experiments~\cite{footnote1}.
Creep of individual microcontacts would therefore be at the nm scale,
which cannot be resolved with our apparatus. 
Particle fracture - which occurs in geophysical experiments and yields 
characteristic grain size distributions - 
plays no role in our studies.
With the ability to image the motion of particles and 
to measure creep and dilation on the ${\rm \mu m}$ scale,
our experiments focus on the role of the arrangement of microcontacts 
(the fabric of the granular material) in determining its shear strength.

\section{Experimental Results} \label{expdata}

\subsection{Apparatus}

\begin{figure}
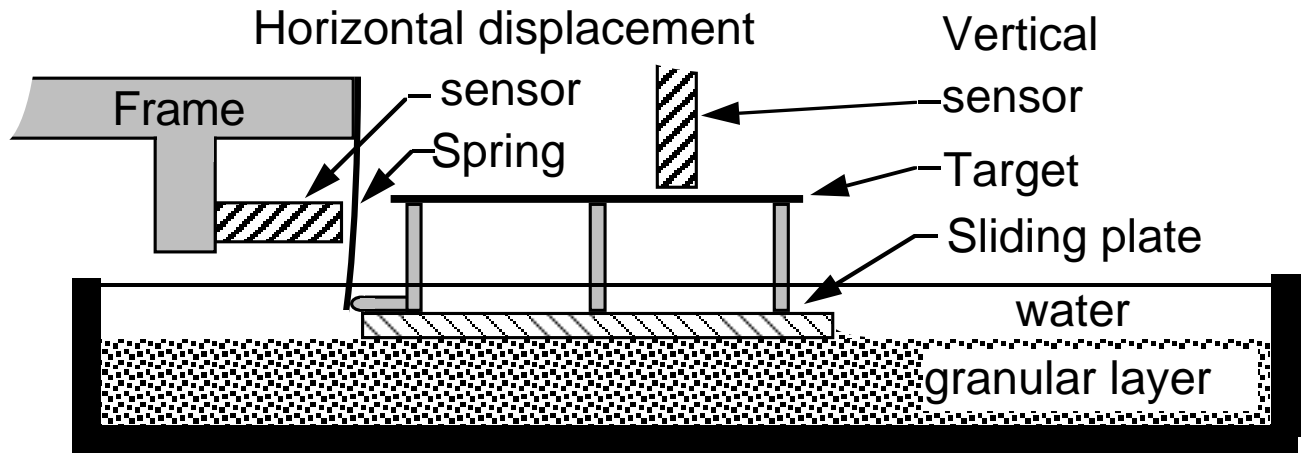
 
\caption{Schematic diagram of the experimental setup for studying shear
in a granular layer with sensitive measurement of horizontal and vertical 
positions under water and in air.}
\label{exp_setup}
\end{figure}

The experimental setup, shown in Figure~\ref{exp_setup} 
has been described previously~\cite{Gollub98},
including the modifications for experiments under water~\cite{Geminard99}.
Our experiments were carried out in a thin $3 {\rm mm}$ layer of 
$103 \pm 14 {\rm \mu m} $ diameter glass beads (Jaygo Inc.) in
a $11 \times 18.5 {\rm cm} $ tray. 
A transparent acrylic plate ($5.28 \times 8.15 {\rm cm}$) 
of weight $26.7$~g is placed on top of the granular material. 
Good contact between plate and layer is assured 
either by etching grooves or gluing a layer of beads to the plate's 
lower surface.
The plate is pushed across the 
granular material by a spring that 
touches only a small steel
ball glued to the plate; this allows for vertical motion of the plate.
The spring's fulcrum is moved toward the plate at constant speed by a microstep
stepper motor.  Bending of the spring is measured with a displacement
sensor, which indicates the force applied to the plate by the spring with
a relative precision of better than $0.1\%$. 
The vertical position of the plate is measured with a second displacement
sensor having a resolution of $\sim 0.1 {\rm \mu m}$.
As indicated in Figure~\ref{exp_setup}, the granular material and the plate 
are kept under water in some experiments.
The motion of particles is imaged from the side with a fast camera 
(Kodak Motioncorder SR-500).

\subsection{Particle motion in sheared granular layers}

In order to establish how deeply the motion of the top plate
penetrates into the granular material, 
we image the position of glass beads at the side of the cell.
A $5.3 \times 18.5 {\rm cm} $ tray - slightly wider than the plate - 
was designed with smooth glass sidewalls to allow for optimal visualization.
Through direct illumination we obtain small circular bright spots
for all beads close to the wall.
We track the motion of all particles in the layer closest to the
wall from images taken at $500$~frames per second during motion of the plate.
(The motion is captured during one slip in the stick-slip regime, which is 
described in section~\ref{strengthening}).
The changes in bead positions between frames are used to calculate
the particle tracks and instantaneous velocities.  
Since the conditions at the wall are not identical 
to the inside of a granular material, motion of particles near the side 
yield only an approximation of particle motion in the interior.
However, the  observations regarding the depth profile
and nonuniformity of motion are unlikely to be qualitatively different
for the interior.

\begin{figure}
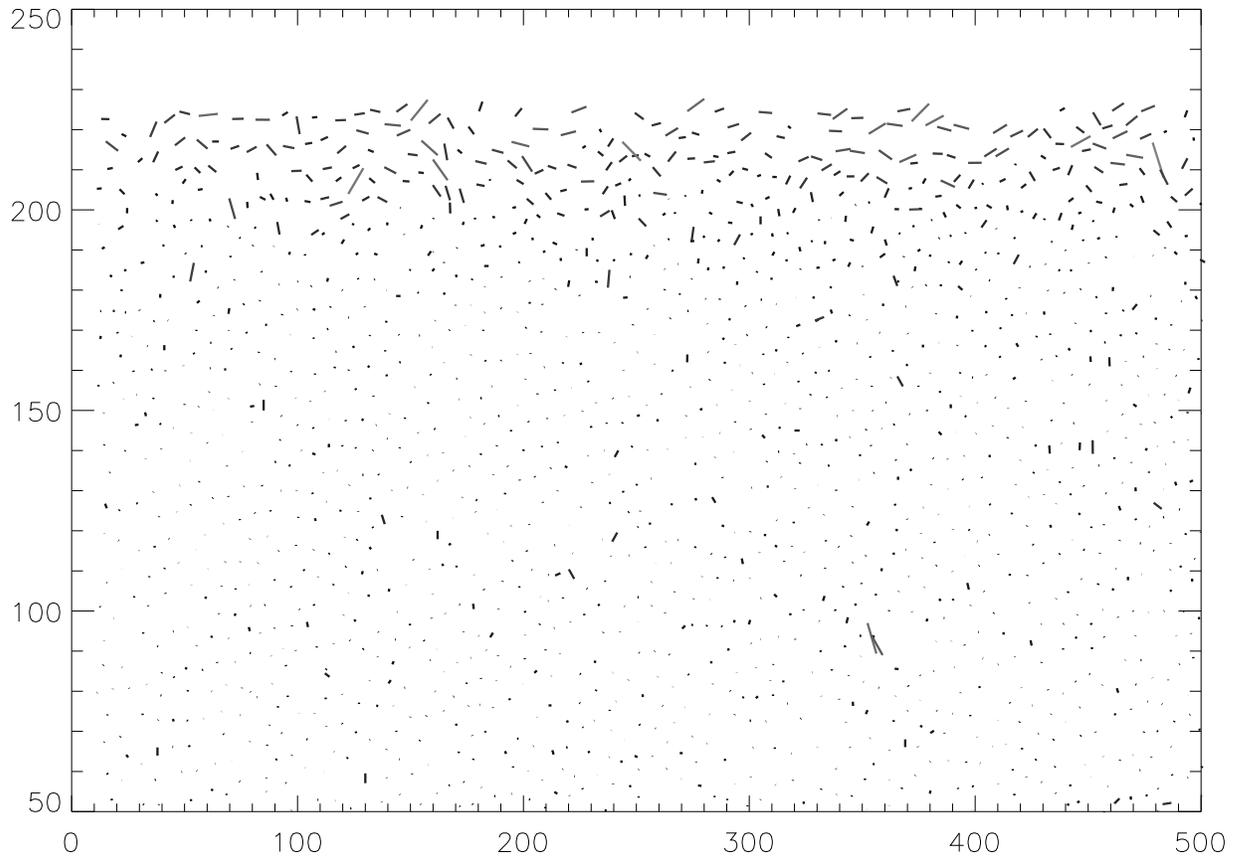
 
\caption{Motion of $\approx 2000$ particles during a short ($40{\rm ms}$) slip.
Particles in approximately 5 layers are moving, but the particle speed
decreases strongly with depth. ($1 {\rm pixel} \simeq 30 {\rm \mu m}$).}
\label{particle_motion}
\end{figure}

One measurement, where $\approx 2000$ glass beads are tracked
during a short ($40 {\rm ms}$) slip, is shown
in Figure~\ref{particle_motion}. Individual particle velocities and directions
of motion are indicated by the direction and length of individual lines. 
Approximately $5$ layers of beads are moving. The particle velocity
decreases more strongly with depth than would be the case for a
fluid, but qualitatively very similar to simulations~\cite{Thompson91}.
The motion of neighboring particles is not perfectly correlated but differs 
in direction and velocity.
This indicates that the material cannot be described as a 
solid with a single fracture plane, 
but that most individual particle contacts are broken up.
Slip of the plate therefore involves the breakup of most particle
contacts within about 5 layers close to the plate: this process
is quite different from 
the breakup of a 2D array of contacts that occurs when 
a solid starts sliding on another solid.  
A more detailed study of particle motion is in progress
and will be reported elsewhere~\cite{Losert99}.

\subsection{Strengthening in dry granular materials}
\label{strengthening}

\begin{figure}
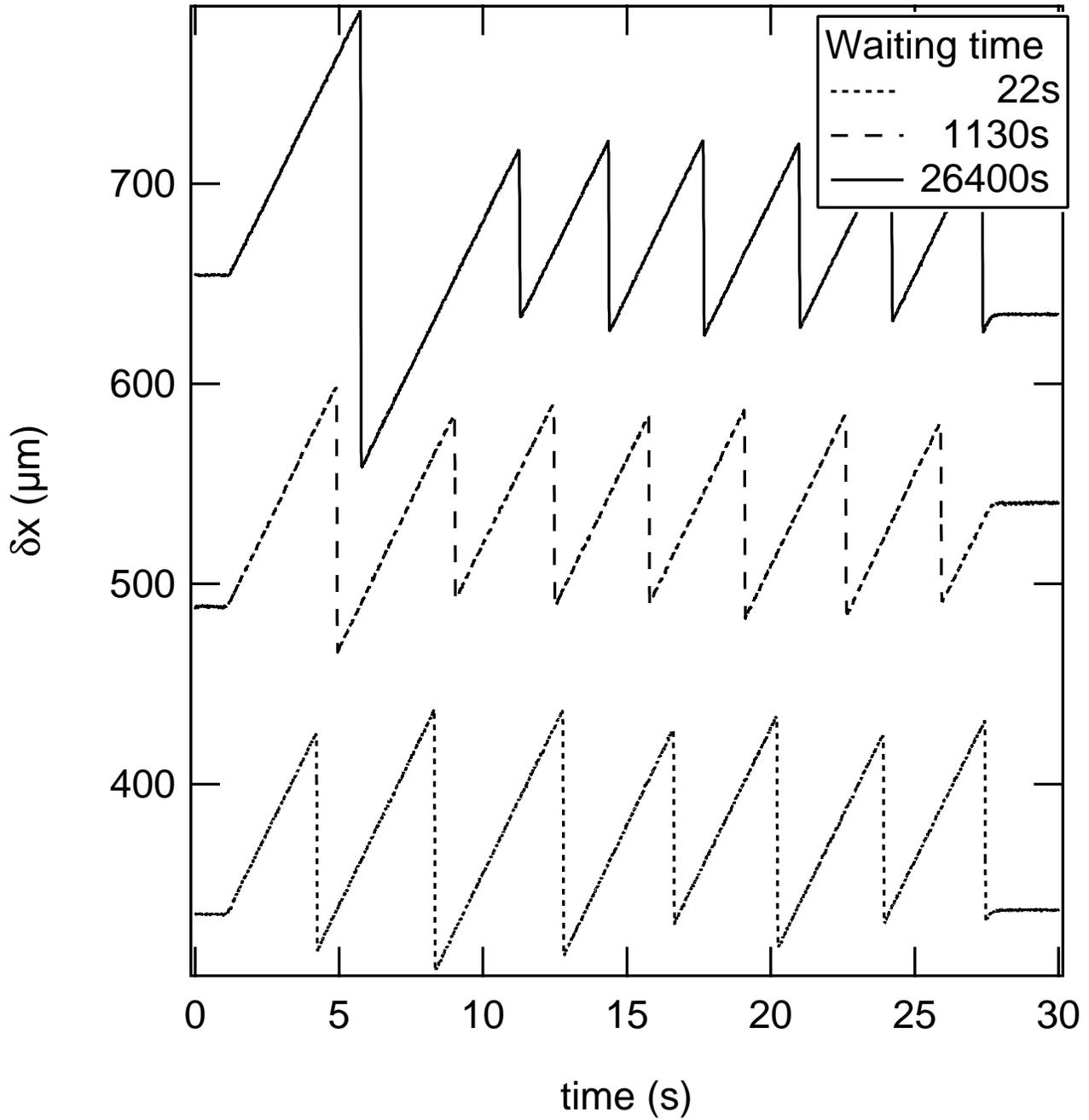
 
\caption{Spring displacement vs time at $v=28.7 {\rm \mu m/s}$,
 $k=189.5 {\rm N/m}$. The motor is stopped during stick-slip motion
and restarted at $t=1.19 {\rm s}$ after a waiting time of $22$~s 
(dotted line), $1130$~s (dashed line, offset by $150 {\rm \mu m}$), 
or $26400$~s (solid line, offset by $300 {\rm \mu m}$).
The maximum spring displacement prior to the first slip 
increases with waiting time. This indicates an increase in the friction
coefficient with the waiting time, when the layer is continuously
held under stress.}
\label{x_vs_t_dry}
\end{figure}

In dry granular materials stick-slip motion with long sticking times 
and short, fast slips is the prevalent behavior at constant 
motor speed in our experiments.  The strength of the material was 
therefore measured from the spring displacement, i.e. the maximum 
spring displacement prior to a slip. In the steady state, the slip reproducibly
starts at the same spring displacement, but the first slip after the motor
is started can be different.
In the first set of experiments to be considered,
the motor is stopped during stick-slip motion
and restarted after a waiting time $\tau$. 
The applied stress during waiting varies by roughly $30\%$ (depending
on how long after a slip event the motor is stopped).
Figure~\ref{x_vs_t_dry}
shows the spring displacement vs time, with the motor started at 
$t=1.19$~s after a waiting time $\tau = 22$~s 
(dotted line), $\tau = 1130$~s (dashed line), or $\tau = 26,400$~s (solid line). 
The total spring displacement before the first slip increases 
with waiting time, but after at most two slips a steady stick-slip motion
is reached with the same maximum spring displacement for all waiting times 
(the curves are offset by $150 {\rm \mu m}$). 

In some experiments no shear stress was applied during the 
waiting time by moving the motor backward prior to waiting. 
The spring displacement when
the motor started with a completely unbent spring at $t=8.2$~s 
after a waiting time of $37,213$~s is shown in 
figure~\ref{d_unstress}. After several slips the motor direction
is reversed at $t=50$~s. In this case, the stress just before
the first slip is not enhanced. At low normal stress 
in the stick-slip regime, the strength of dry 
granular material therefore
only increases with waiting time if a shear stress is applied during the 
waiting time.  

\begin{figure}
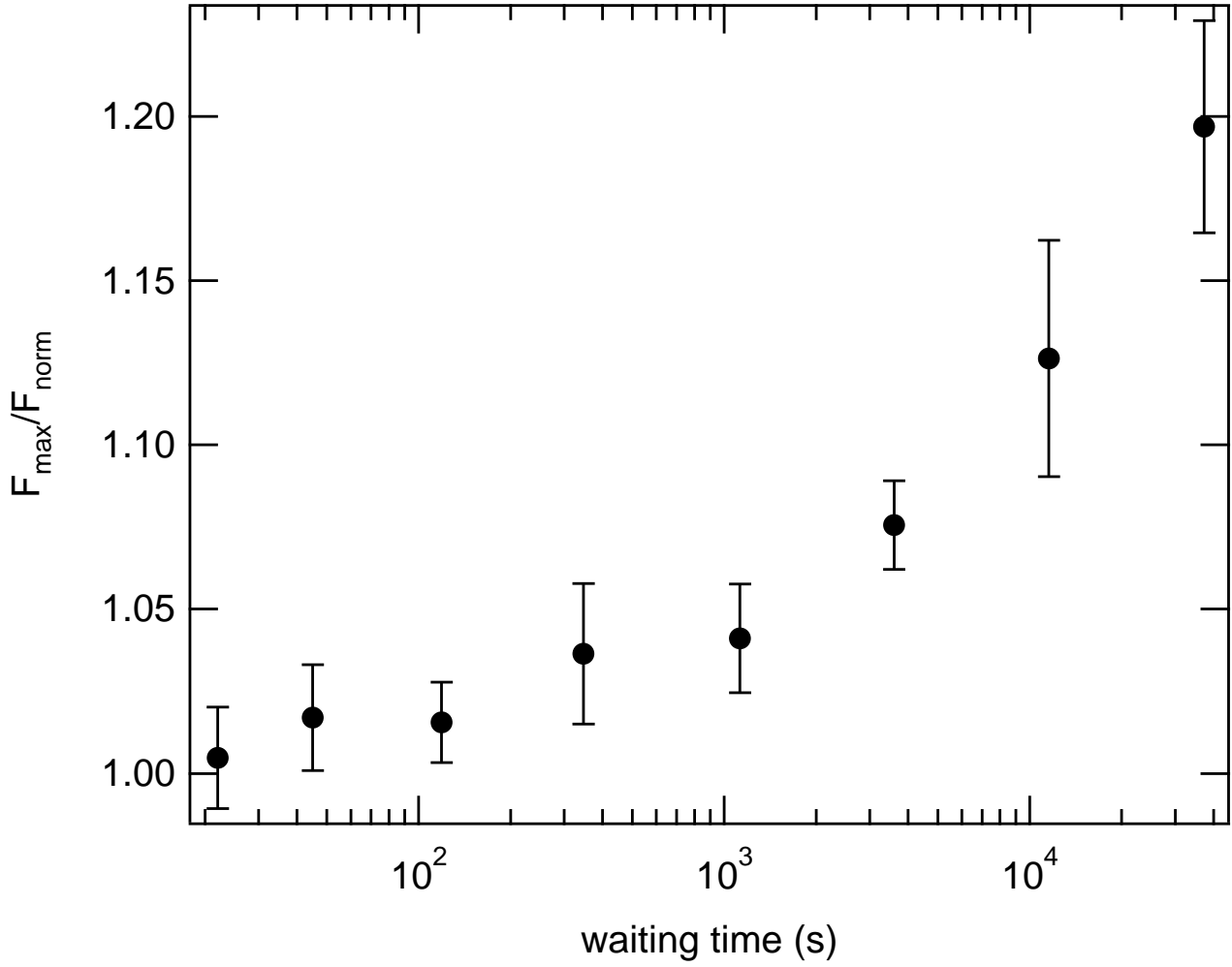
 
\caption{Strengthening: Ratio of maximum static friction 
force $F_{max}$ after a waiting time compared to the maximum static 
friction force $F_{norm}$ during continuous 
stick-slip motion at $v=28.7 {\rm \mu m/s}$, $k=189.5 {\rm N/m}$. 
The strength increases roughly logarithmically with 
waiting time for longer times. Each point is an average over several runs.}
\label{strength_vs_wait}
\end{figure}

The ratio of the maximum spring displacement prior to the first slip $F_{max}$
to the average maximum spring displacement $F_{norm}$ 
of all except the first two slips
indicates the relative strengthening of the material with waiting time.
This ratio is shown in figure~\ref{strength_vs_wait} for the
stressed case.  
The waiting time strengthening under shear stress
appears to be faster than logarithmic when the full range of waiting
times $\tau$ is included. 
On shorter timescales the material becomes roughly $2\%$ stronger (compared
to continuous stick-slip motion) per decade increase in $\tau$.
On longer timescales $\tau > 1000 $~s the 
strengthening is roughly $10\%$ per decade 
with a characteristic initiation time
$\tau_0$ (the time before which little
strengthening occurs based on an extrapolation
of the approximately logarithmic increase) of roughly $600 \pm 300$~s.

\subsection{Role of Dilation}

The vertical position $h(t)$ of the plate indicates the dilation or 
compaction of the granular material.
Due to small deviations from perfect flatness of the
target plate for the vertical displacement sensor and 
slow fluctuations in the sensor readout due to
small variations in temperature, the absolute dilation cannot be
computed to better than $1 {\rm \mu m}$ 
over long times or over horizontal plate movements 
long compared to the particle diameter. 
 
\begin{figure}
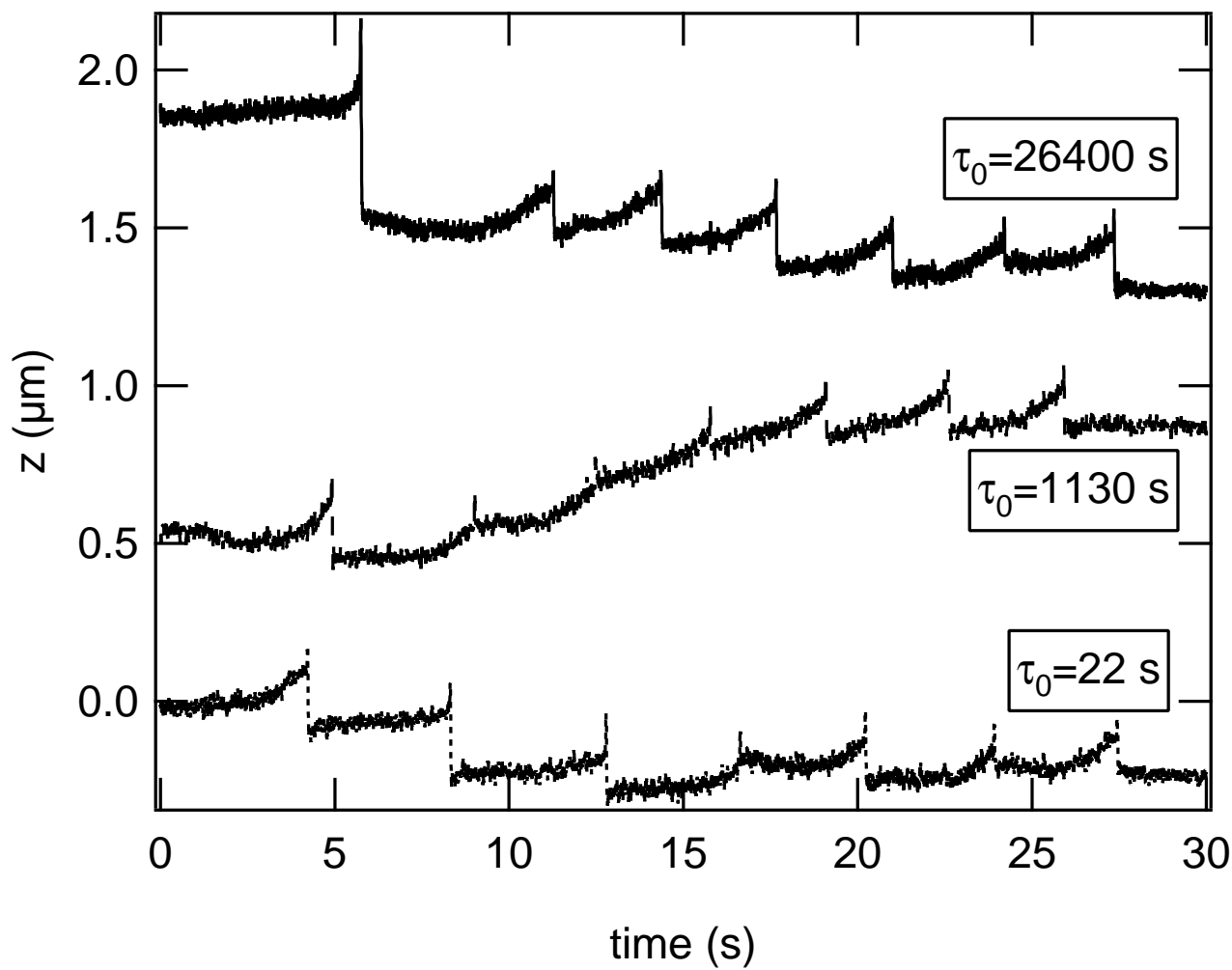
 
\caption{Dilation during stick-slip motion (dry).
Vertical plate displacement vs time at $v=28.7 {\rm \mu m/s}$,
 $k=189.5 {\rm N/m}$. The motor is stopped during stick-slip motion
and restarted at $t=1.19 {\rm s}$ after a waiting time of $22$~s 
(dotted line), $1130$~s (dashed line), or $26400$~s (solid line).
The compaction following the first slip increases with waiting time;
this indicates that dilation occurred  during the waiting time.}
\label{z_vs_t_dry}
\end{figure}

Figure~\ref{z_vs_t_dry} shows the vertical 
plate position after a waiting time under shear stress
for the experiments described in Figure~\ref{x_vs_t_dry}. 
After the motor is started, 
a gradual dilation of the granular material during the sticking time
takes place, followed by rapid dilation as the slip starts and 
compaction immediately following the slip. The first slip after a
long waiting time (solid line) is followed by especially strong compaction.

\begin{figure}
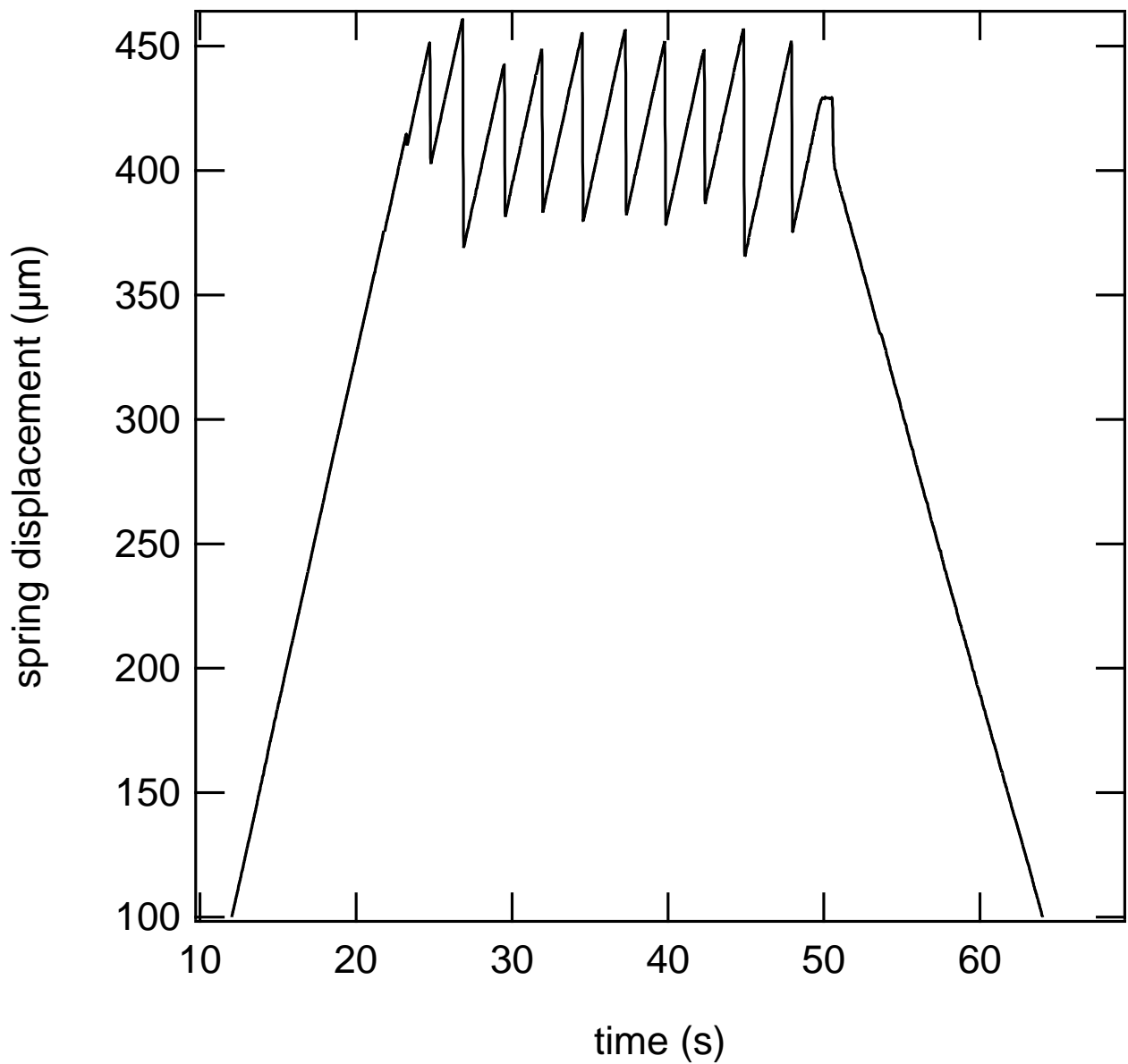
 
\caption{Spring displacement vs time without initial applied stress 
($v=28.7 {\rm \mu m/s}$, $k=189.5 {\rm N/m}$). The motor 
started with a completely unbent spring at $t=8.2$~s after 
a waiting time of $37213$~s (solid line). The motor is reversed 
at $t=50$~s.  The maximum spring displacement
does not change with waiting time.}
\label{d_unstress}
\end{figure}

\begin{figure}
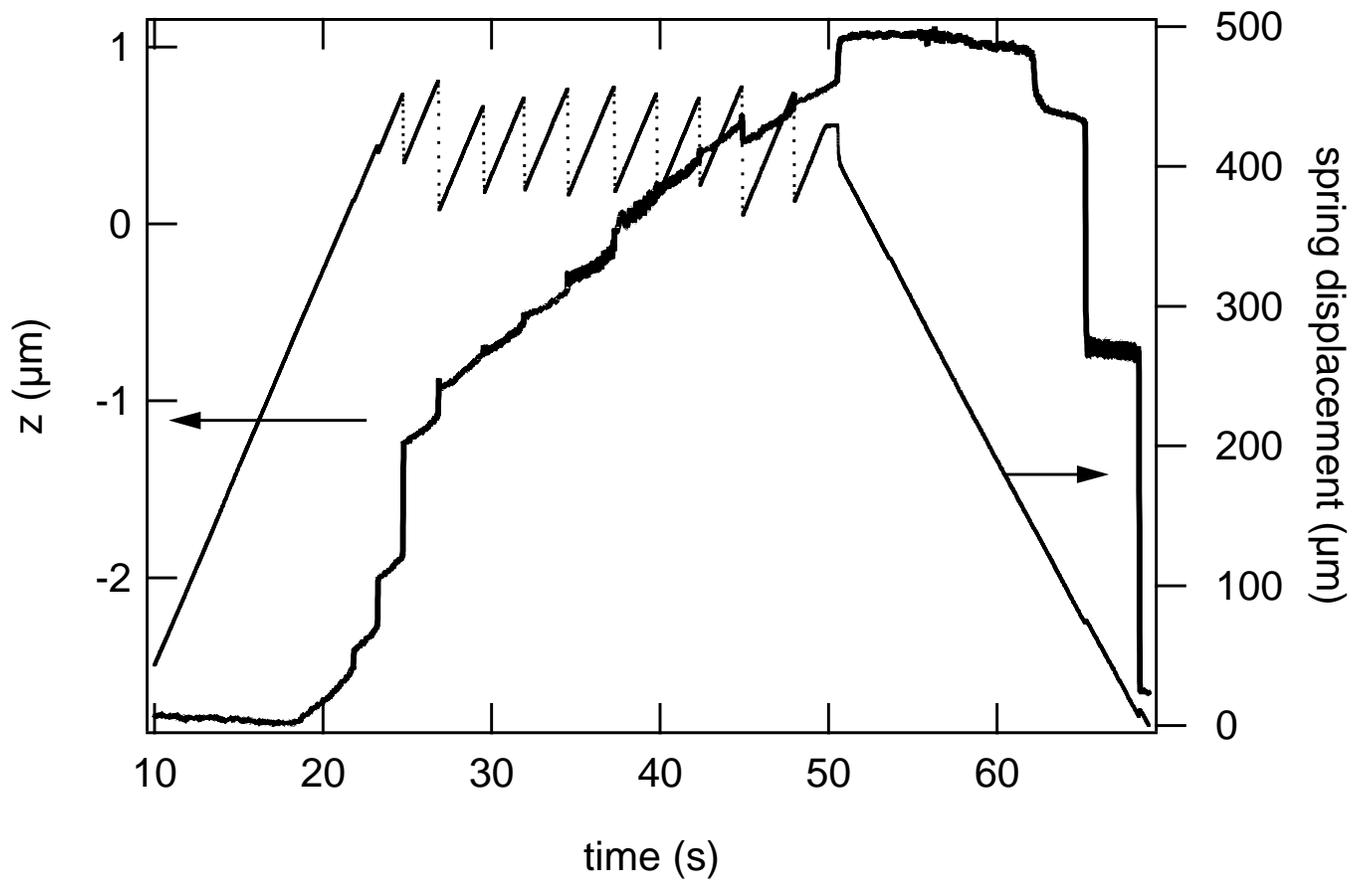
 
\caption{Vertical displacement vs time (solid line)  
and spring displacement (dotted line) for the data of Figure~4
($v=28.7 {\rm \mu m/s}$, $k=189.5 {\rm N/m}$). The motor 
started with a completely unbent spring at $t=8.2$~s after 
a waiting time of $37213$~s. The material dilates when motion
starts.  
The reduction of the applied stress is accompanied by compaction.}
\label{z_unstress}
\end{figure}

The vertical dilation {\it without} applied shear stress during waiting
is shown in Figure~\ref{z_unstress}, which corresponds to the spring
displacement data of Figure~\ref{d_unstress}.
As the plate starts to move, the layer gradually dilates. The
gradual dilation does not influence the maximum frictional force,
which is comparable for all slips as shown in Figure~\ref{d_unstress}.
As soon as the shear stress is released at $t > 60$~s, 
the layer compacts almost instantaneously.

\subsection{Strengthening under water}

\begin{figure}
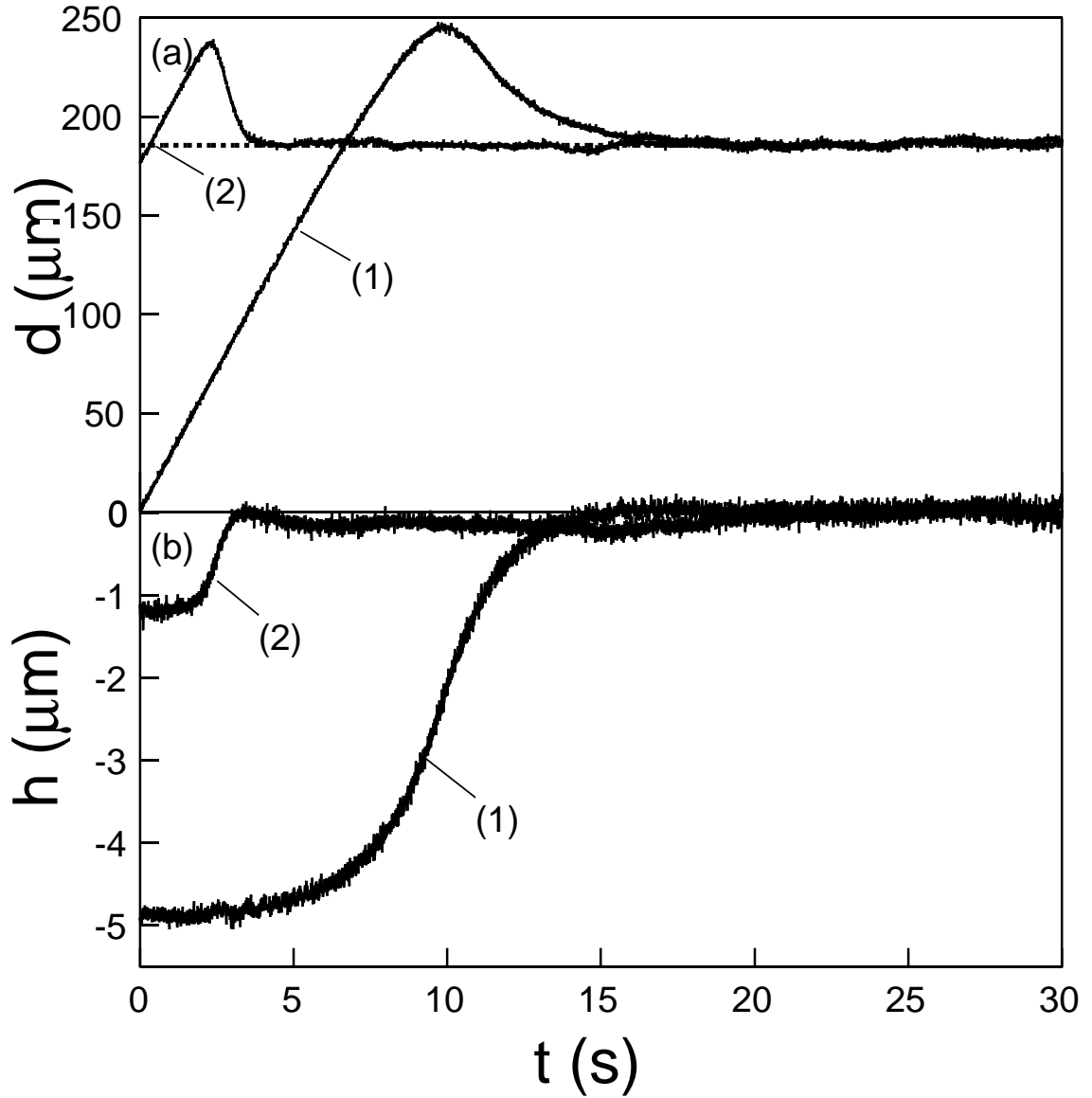
 
\caption{(Wet friction) Behavior (a) of the spring displacement $d(t)$ and
(b) of the vertical position $h(t)$ as functions
of time $t$ in two different cases: (1) The horizontal stress
is released before the experiment;
(2) The horizontal stress is continuously applied.
In the second case, the layer is initially less packed because of the
applied shear stress; 
as a consequence, the total dilation $\Delta h$ observed during 
the experiment is less
($k = 189.5~\rm N/m$, $M = 14.5~{\rm g}$, $V = 28.17~{\rm \mu m/s}$)
(from Ref.[6]). }
\label{sample_result_wet}
\end{figure}

Since the formation of liquid bridges between particles 
could be the cause for the
observed strengthening of granular material, as described 
in~\cite{Ciliberto98}, a second set of experiments was carried out 
under water, where liquid bridges are absent.
Continuous sliding is prevalent under water because the fluid lubricates the
contacts. Figure~\ref{sample_result_wet}
shows  the typical behavior  of the spring displacement $d(t)$
(Fig.~\ref{sample_result_wet}a) and
 of the vertical position $h(t)$ (Fig.~\ref{sample_result_wet}b) 
as functions
of time $t$ in two different cases: (1) The horizontal stress
is released before the experiment;
(2) The horizontal stress is continuously applied.
In both cases the  spring displacement  during
the transient (which is proportional to
the frictional force at the small accelerations considered here) 
is larger than the frictional force
during steady sliding. The force reaches a maximum
$F_{max}$ at the maximum dilation rate. At later times
the material continues to dilate and approaches a steady
state dilation, while the frictional force decreases toward a steady sliding 
frictional force.
In case (2), the layer is initially less packed and
the total dilation $\Delta h$ observed during 
the experiment is smaller. 

As observed for the dry granular material, we find that the 
maximum frictional force $F_{max}$ depends
on the resting time $\tau$, and on the horizontal stress
applied during this interval, as shown in Figure~\ref{wetstrengthening}.
The experimental procedure is as follows:
The plate is initially pushed at constant velocity
($28.17 \mu m/s$) until the steady state regime is reached
with a dynamic friction force $F_d$. 
Then, the motion of the translator is suddenly stopped and
the plate stops at a well-defined horizontal applied stress
($F = 3.2~10^{-2} N \simeq F_d$)~\cite{footnote2}.
The translator motion is started again after a delay $\tau$.
As for dry granular materials,
we find that the maximum value of the frictional force $F_{max}$
depends on the waiting-time $\tau$:
The maximum frictional force increases by roughly $10\%$ of $F_d$
for each order of magnitude increase in waiting time with a characteristic
initiation time $\tau_0 = 0.6 \pm 0.2$~s, a factor of $\sim 10^3$ faster 
than for a dry granular material.
The maximum frictional force $F_{max}$ increases by about $40\%$ in
$10$ hours. 
On the other hand, if no horizontal stress is applied during the
waiting-time $\tau$ (the spring is pulled back), no
increase of the maximum frictional force is measured.
For short waiting times $F_{max}$ is found to be larger when no
stress is applied;
this result is in agreement with the fact that the 
maximum frictional force increases with the total dilation
of the layer $\Delta h$ in the continuous sliding case for a wet granular 
material~\cite{Geminard99}. 
The two curves for strengthening with and without applied shear stress
intersect for $\tau \simeq 10^4 $~s. For longer waiting times
$F_{max}$ becomes larger for waiting under an applied shear stress
while the dilation following the waiting ($\Delta h$) remains larger
when no stress is applied. 

\begin{figure}
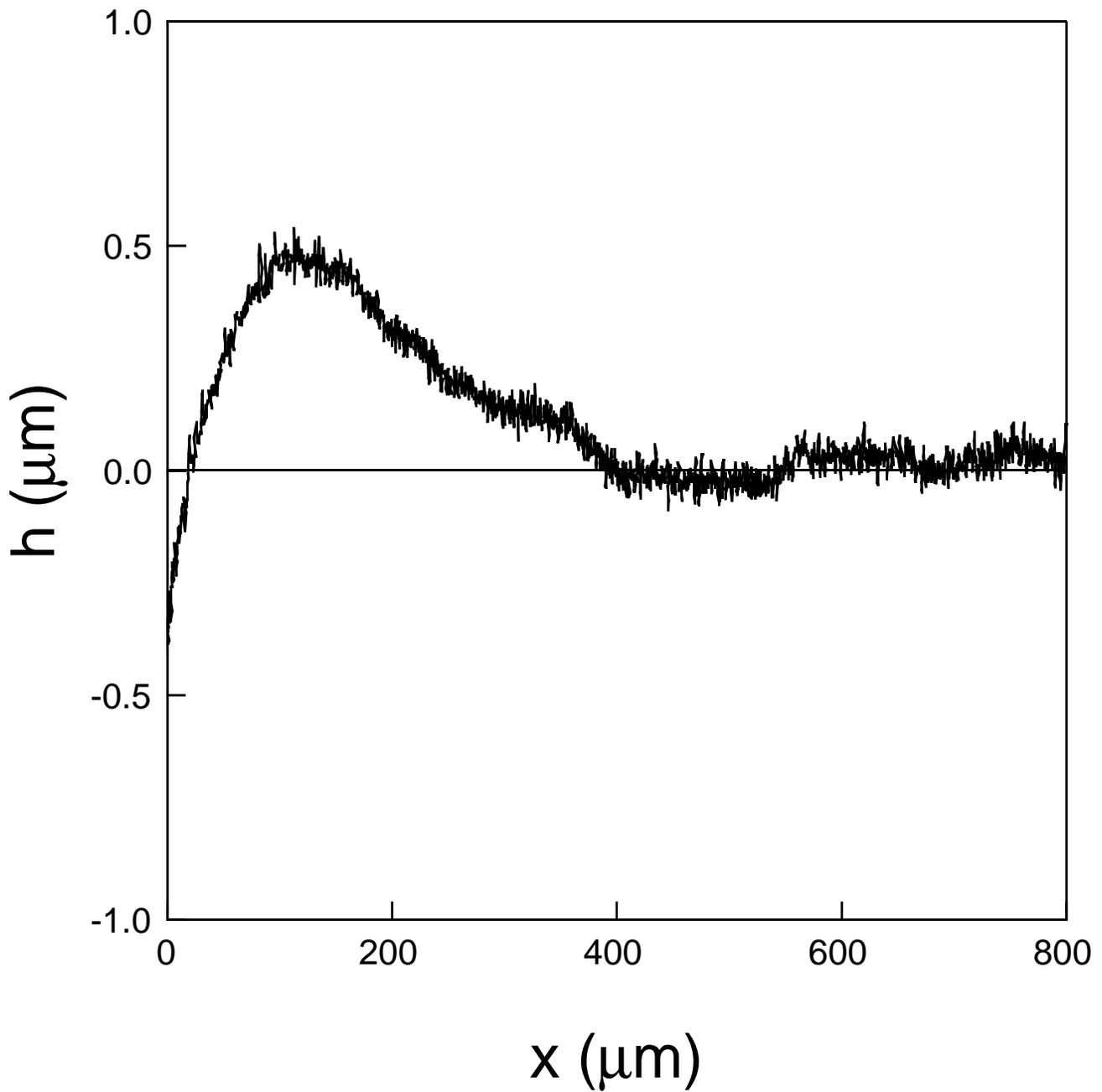
 
\caption{(Wet friction) 
Vertical position of the plate $h$ as a function of its
horizontal position $x$ during a transient.
The vertical position overshoots when 
the plate was subjected to a horizontal
stress during the waiting time $\tau$
($k = 189.5 {\rm N/m}$, $M = 14.5 {\rm g}$, $V = 28.17 {\rm \mu m/s}$,
$\tau = 26000 {\rm s}$).}
\label{h_vs_x_wt}
\end{figure}

	A careful study of the behavior of the vertical
position of the plate $h$ as a function of its horizontal
position $x$ shown in Figure~\ref{h_vs_x_wt} indicates
that $h$ reaches a maximum during
the transient regime when a horizontal stress is applied during the waiting 
time.
The distance over which the
vertical position of the plate reaches its maximum is about 
one particle radius $R$, comparable to
the distance over which the system reaches the steady state regime
in experiments starting without applied shear stress~\cite{Geminard99}. 
However, after a waiting time under stress, the vertical position of the plate
decreases for a sliding distance of a few particle diameters.
The frictional force reaches its asymptotic value over the
distance $R$, and appears not to be affected by this decrease
of the vertical position of the plate. 

Creep during the waiting time can be directly observed. 
If the plate is kept under a shear stress comparable 
to the steady sliding value,
it creeps slowly (${dx/dt} \simeq 1{\rm \mu m/h}$)
and goes up (Fig.\ref{waiting_time}).  Such strong creep has not been observed 
in dry granular materials.
The vertical velocity $dh/dt$ of the plate is of the same
order of magnitude as the horizontal velocity $dx/dt$.
Moreover, the whole vertical displacement of the plate $\delta h$
after $7$ hours is about $20 \mu m$, much larger than any vertical
displacement of the plate observed in the dynamical regime.
 
\section{Discussion and Conclusion} \label{discussion}

\begin{figure}
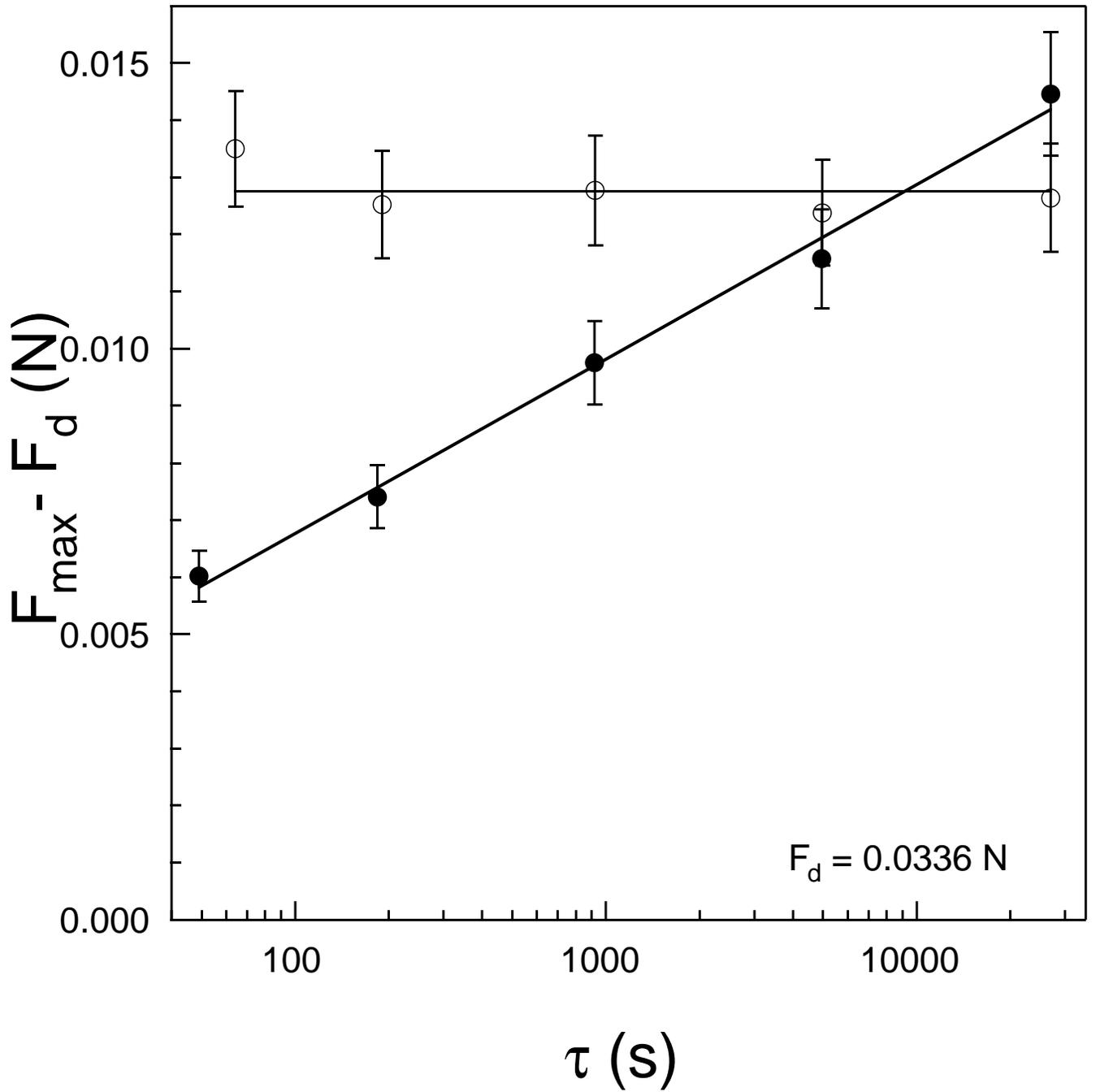
 
\caption{(Wet friction) Maximum frictional force $F_{max}$ as a function 
of the waiting time $\tau$
($k = 189.5 {\rm N/m}$, $M = 14.5 {\rm g}$, $V = 28.17 {\rm \mu m/s}$).
Empty circles: no horizontal stress
is applied during the waiting time $\tau$. Filled circles: the plate
is submitted to a horizontal stress ($F \simeq F_d$)
during the waiting-time $\tau$.}
\label{wetstrengthening}
\end{figure}

The main results of this investigation of the shear strength of
a granular material  at low normal forces are as follows:

(a) The strength of a granular material increases roughly logarithmically 
with the time of stationary contact $\tau$ (waiting time)
in both dry material (Fig.~\ref{strength_vs_wait}) and wet material
(Fig.~\ref{wetstrengthening}), 
if a shear stress is applied during the waiting time.
The characteristic initiation time $\tau_0$ is roughly three 
orders of magnitude
smaller for wet granular material.

(b) In both dry and wet granular matter, the strength 
does not increase with $\tau$ if no shear stress
is applied during the waiting time (Figs.~\ref{d_unstress} and 
\ref{wetstrengthening}).

(c) In both the dry and the wet granular material,
the layer compacts immediately, when the shear
stress is released. This compaction leads to an
instantaneous increase in the strength of the wet
granular material, but not in the dry case (Figs.~\ref{d_unstress} and 
\ref{wetstrengthening}).

(d) Particle tracking reveals that
in a region of approximately $5$ particle layers below the sheared
surface, particles move and lose contact with each other during
stick-slip motion (Fig.~\ref{particle_motion}).
The particle velocities within this fluidized region decrease faster than 
linearly. 

(e) Dilation or compaction often influences the strength of the
material. However, 
gradual slight dilation in a dry material (Fig.~\ref{z_unstress})
and gradual slight  compaction in a wet material
(Fig.~\ref{h_vs_x_wt}) do not necessarily influence the frictional forces.

In order to understand these results we need to
go back to the main question: 
What are the microscopic mechanisms for strengthening?
In the following paragraphs we look at individual mechanisms
and the experimental results that support them or show their limitations.

The gradual strengthening of individual microcontacts through
creep on the level of microcontacts has been found 
in other work to lead to
strengthening in solid-on-solid friction.
The microcontact size is determined by
the yield stress of the material. Unless the contact area is flat,
the material deforms until the microcontact is just large enough to support 
the stress. Strengthening of microcontacts should therefore be present 
for granular materials with
nm scale contact areas just as it is for solid-on-solid friction.  

Another mechanism that can strengthen microcontacts is 
an attractive surface 
force, created by gradually developing liquid bridges between 
particles at rest.
This mechanism  accounts for the observed
increase in strengthening at higher
humidity, but can not explain why strengthening also occurs for
a granular layer under water. Other mechanisms for strengthening
must therefore be present.

No strengthening is observed in the absence of
an applied shear stress even though the absence of applied shear stress
only decreases the total stress on microcontacts by $10-20\%$.
In addition, 
the observed creep distance under water (Fig.~\ref{waiting_time})
is several orders of magnitude larger than the characteristic
microcontact length; this suggests that microcontacts break 
rather than strengthen during the time at rest.
Microcontact strength can therefore not account for strengthening under water
or the lack of strengthening in the absence of an applied shear stress
in both dry and wet material.

The large creep under water suggests that grains rearrange more easily 
in wet than
in dry granular matter. One possible explanation is that water lubricates 
individual contacts between grains, which allows grains to slip past each
other with significantly smaller frictional force. Lubrication
is also evident in the reduced friction coefficient of a wet granular 
material~\cite{Geminard99}.   
Even though large creep decreases the contact time of microcontacts, 
strengthening occurs at the same rate, and 
with a much smaller characteristic 
initiation time $\tau_0$ in wet granular matter.
This indicates that rearrangments in the network of contacts 
contibute significantly to strengthening. 
The rigidity of the network of contacts 
can influence the frictional strength of the material,
since direct observations indicate that a
three dimensional network of contacts is broken (and a fluidized region 
of several layers close to the sheared surface forms) when sliding starts.

The observation of strengthening solely under shear stress
differs from results in solid-on-solid friction, but is similar to
results at geophysical pressures~\cite{Marone98b,Marone98a}.
It implies that the contact network changes in response
to an applied shear stress regardless of how the contacts
themselves evolve. 
This is consistent with the description of a granular material 
as a fragile network that breaks if the direction of the applied
stress changes. Recent experiments in a Couette cell, which will
be reported in detail elsewhere~\cite{Losert99b}, support the 
conclusion that the direction of the applied stress matters. 
In the Couette cell a clockwise shear stress that is less than the 
slip threshold is first applied for a waiting time $\tau$.
If the granular material is subsequently sheared clockwise,
a roughly logarithmic increase in the strength with $\tau$ is observed, 
as for the flat layer studied in this paper.
On the other hand, if the subsequent shear is counterclockwise,
no strengthening is observed. 

The density of the granular material is a rough indicator of the
density of the contact network, as more grains tend to touch each
other in denser configurations.
However, compaction and dilation are not always correlated with 
frictional strength in our experiments, possibly because we are limited to
measuring the mean density, not the local density of the surface region,
where the network of microcontacts fails. The possibility that
the orientation of microcontacts could be as important for the strength of 
the material as the density of contacts should also be explored further.

When the shear stress is completely released,
a significant density change occurs. 
This suggests that the granular material under normal stress can only
compact if the existing contact network breaks - which in our case 
happens when the shear stress is removed.
Recent experiments in a Couette cell~\cite{Losert99b}, 
show that changing the direction of shear stress 
can be used to strengthen the granular material rapidly.
The inner cylinder of the Couette cell is 
connected to a motor through a soft spring, which allows for variations
in the applied shear force at forces below the frictional strength of the
granular material.  
When the direction of the applied shear stress is reversed periodically
at stresses below that necessary to initiate sliding, the strength of
the material can be increased rapidly in proportion to
the number of direction reversals.

In conclusion, 
strengthening in these experiments can be
explained by two fundamentally different phenomena. One is a strengthening
of individual microcontacts due to time of contact or slow creep - or due
to the formation of liquid bridges.
The other fundamental strengthening mechanism is related to the spatial
arrangement of beads and hence the arrangement and orientation of
microcontacts.  The strength of the contact network (sometimes called the
fabric of the granular material) 
can be related to the compaction of the granular material,
but our experiments indicate that compaction is not the only
determinant of the strength of the network.

Strengthening due to rearrangements alone can also be related to 
a wide range of other systems that can jam, such as molecules in a glass. 
It has recently been suggested~\cite{Liu98,Edwards99} 
that some aspects of the jamming behavior,
or the 'unjamming' which we study, might have a common theoretical 
description, so a good understanding of the strengthening 
might eventually be useful for a description of the properties of these other
systems as well.

\section{Acknowledgments}
This work was supported by the U.S. National Science Foundation under Grant 
DMR-9704301. 
J.-C. G. thanks the Centre National de la Recherche Scientifique (France) 
for supporting the 
research of its members carried out in foreign laboratories. 
The optical measurements were carried out by P. Ingebretson. 
We appreciate helpful discussions with C. Scholz and C. Marone.

\newpage

\begin{figure}
\caption{(Wet friction)
Horizontal and vertical positions of the plate $x$ 
and $h$ as functions of time $t$ when the system is submitted
to a static horizontal applied stress $F \simeq F_d$. Creep of
about $1{\rm \mu m/h}$ is clearly evident
($k = 189.5 {\rm N/m}$, $M = 14.5 {\rm g}$).}
\label{waiting_time}
\end{figure}

\setcounter{figure}{0}
\pagebreak 

\begin{figure} \begin{center}
\epsfig{file=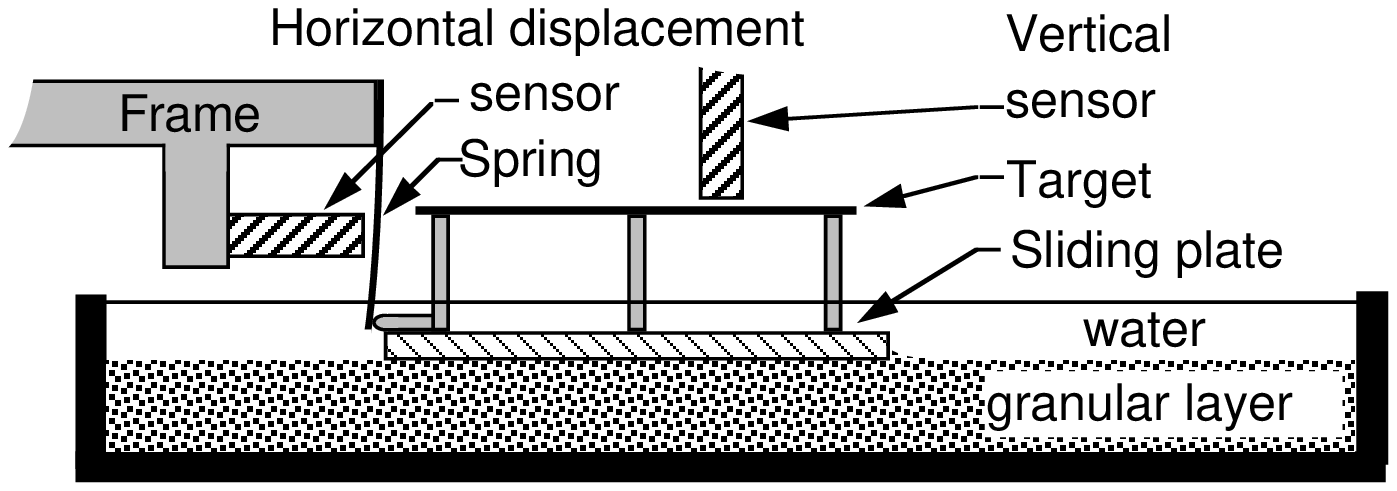,width=\linewidth}
\end{center}
\caption{Schematic diagram of the experimental setup for studying shear
in a granular layer with sensitive measurement of horizontal and vertical 
positions under water and in air.}
\end{figure}

\begin{figure} \begin{center}
\epsfig{file=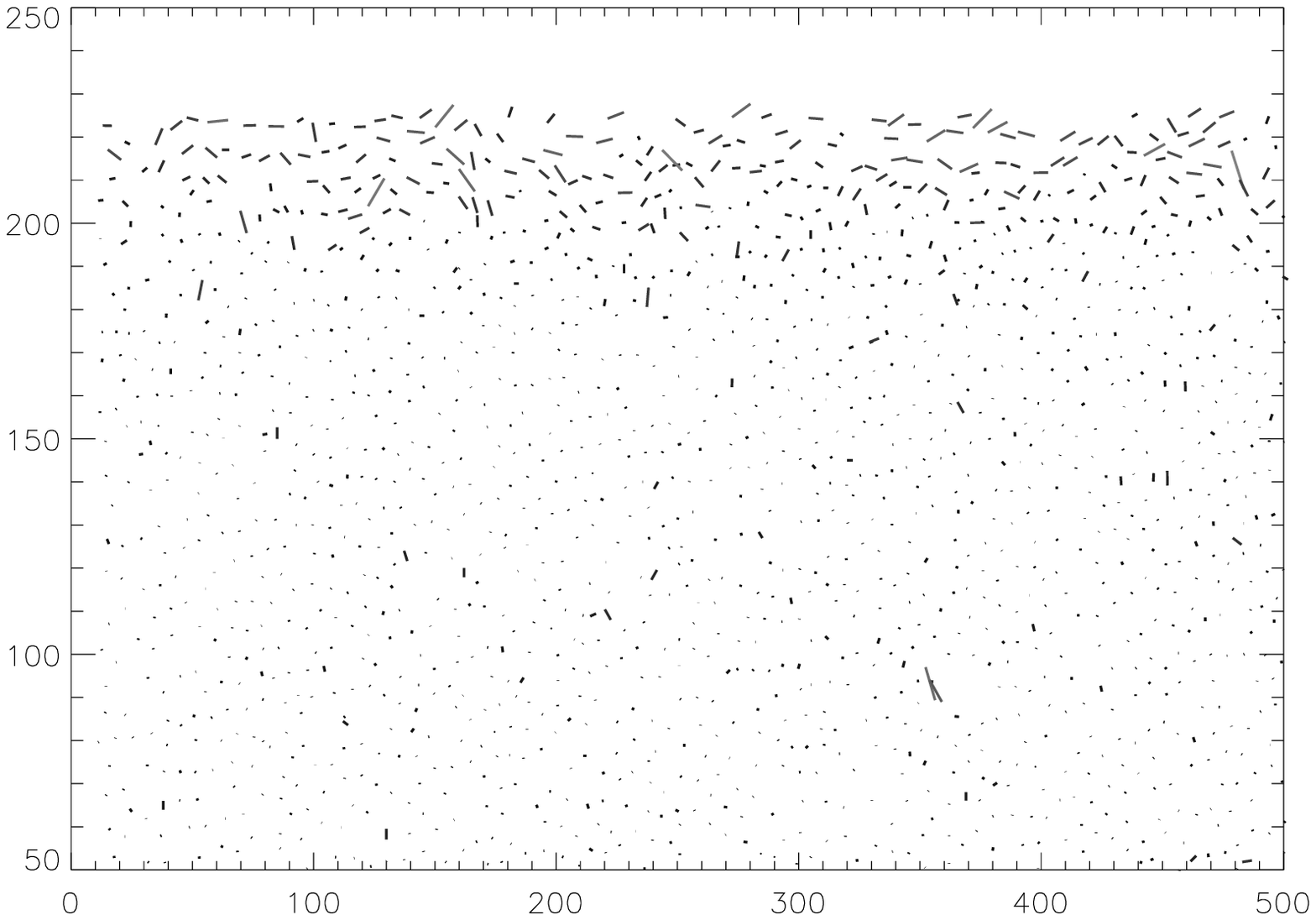,width=\linewidth}
\end{center}
\caption{Motion of $\approx 2000$ particles during a short ($40{\rm ms}$) slip.
Particles in approximately 5 layers are moving, but the particle speed
decreases strongly with depth. ($1 {\rm pixel} \simeq 30 {\rm \mu m}$).}
\end{figure}

\begin{figure} \begin{center}
\epsfig{file=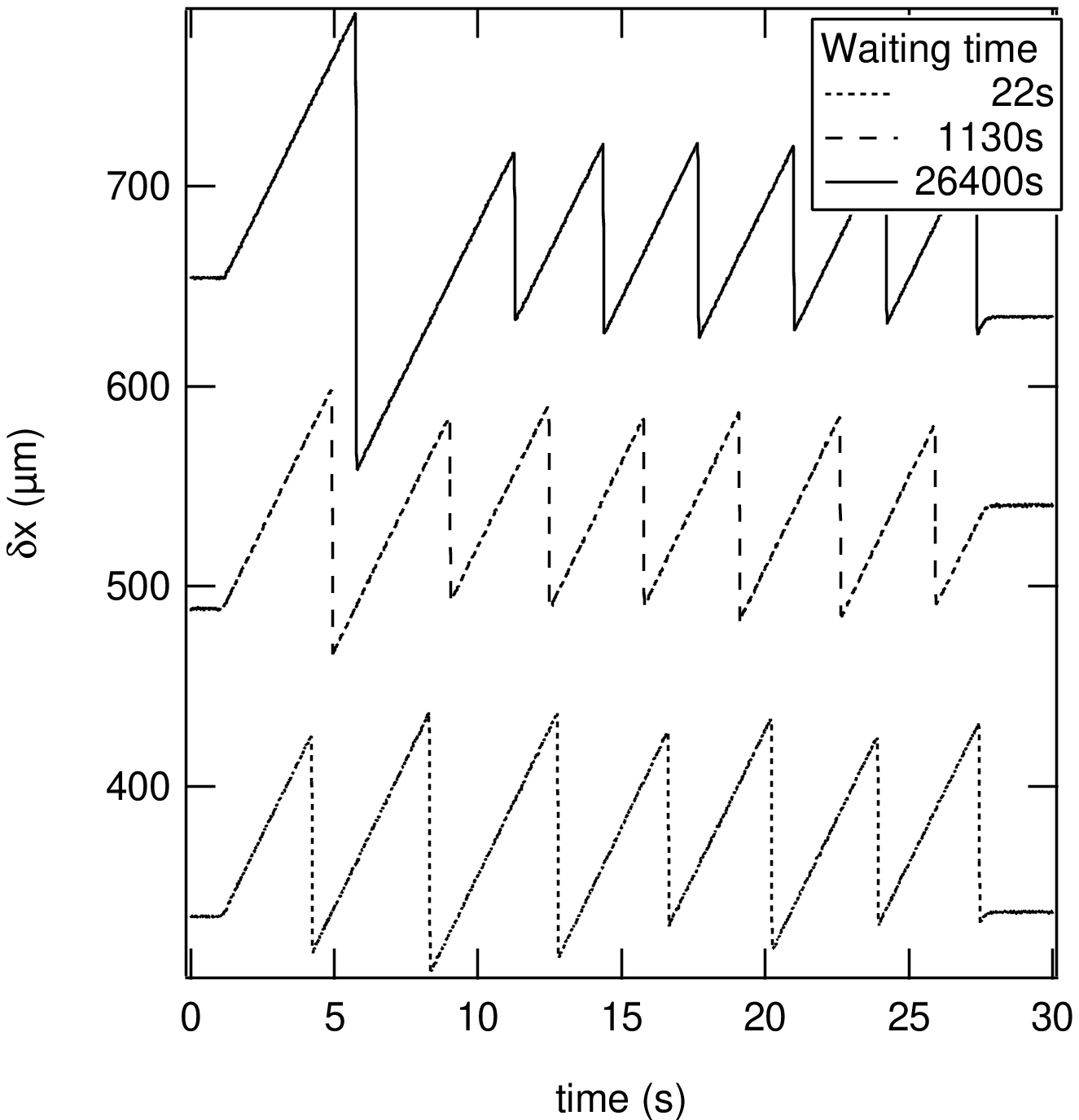,width=\linewidth}
\end{center}
\caption{Spring displacement vs time at $v=28.7 {\rm \mu m/s}$,
 $k=189.5 {\rm N/m}$. The motor is stopped during stick-slip motion
and restarted at $t=1.19 {\rm s}$ after a waiting time of $22$~s 
(dotted line), $1130$~s (dashed line, offset by $150 {\rm \mu m}$), 
or $26400$~s (solid line, offset by $300 {\rm \mu m}$).
The maximum spring displacement prior to the first slip 
increases with waiting time. This indicates an increase in the friction
coefficient with the waiting time, when the layer is continuously
held under stress.}
\end{figure}

\begin{figure} \begin{center}
\epsfig{file=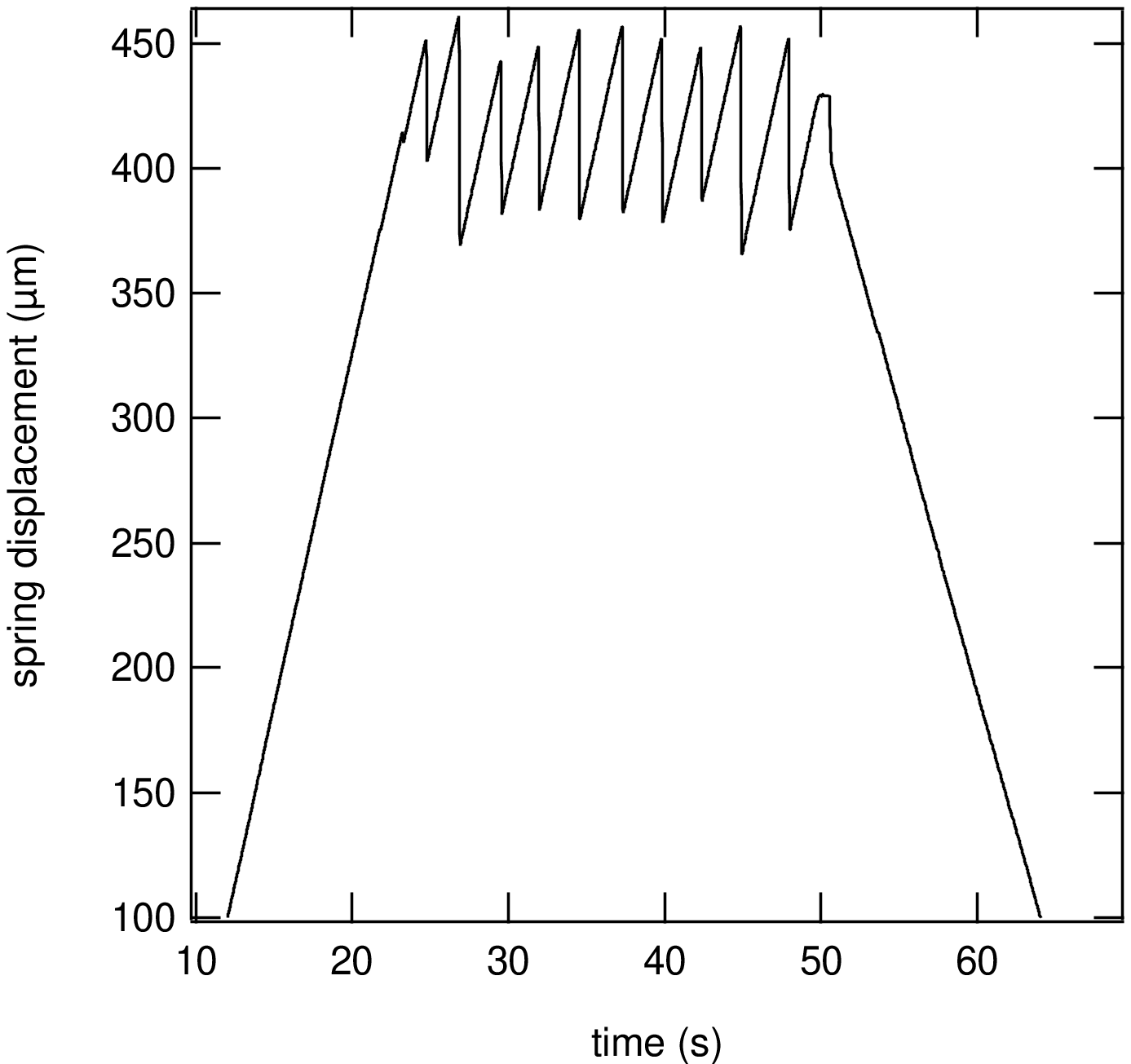,width=\linewidth}
\end{center}
\caption{Spring displacement vs time without initial applied stress 
($v=28.7 {\rm \mu m/s}$, $k=189.5 {\rm N/m}$). The motor 
started with a completely unbent spring at $t=8.2$~s after 
a waiting time of $37213$~s (solid line). The motor is reversed 
at $t=50$~s.  The maximum spring displacement
does not change with waiting time.}
\end{figure}

\begin{figure} \begin{center}
\epsfig{file=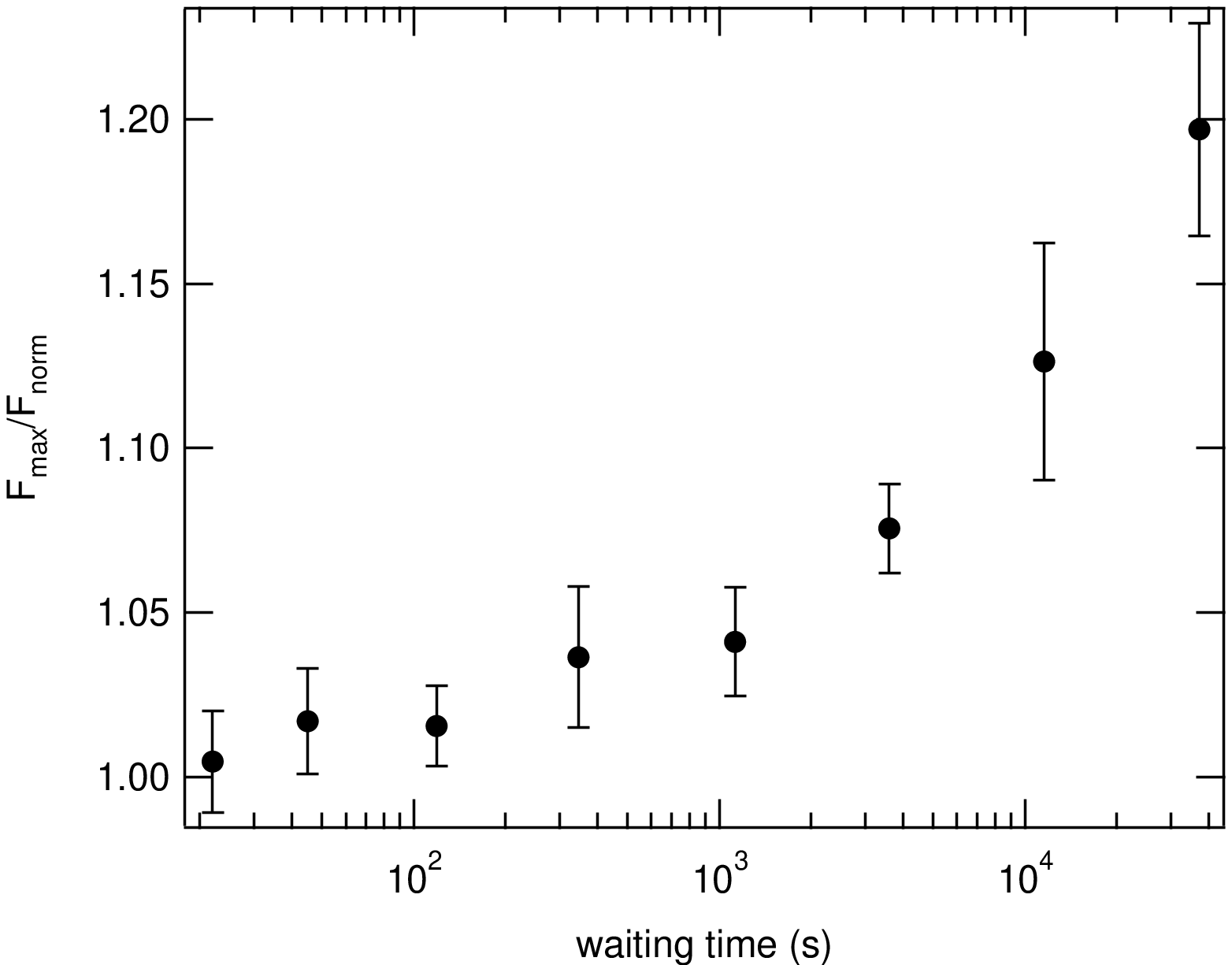,width=\linewidth}
\end{center}
\caption{Strengthening: Ratio of maximum static friction 
force $F_{max}$ after a waiting time compared to the maximum static 
friction force $F_{norm}$ during continuous 
stick-slip motion at $v=28.7 {\rm \mu m/s}$, $k=189.5 {\rm N/m}$. 
The strength increases roughly logarithmically with 
waiting time for longer times. Each point is an average over several runs.}
\end{figure}

\begin{figure} \begin{center}
\epsfig{file=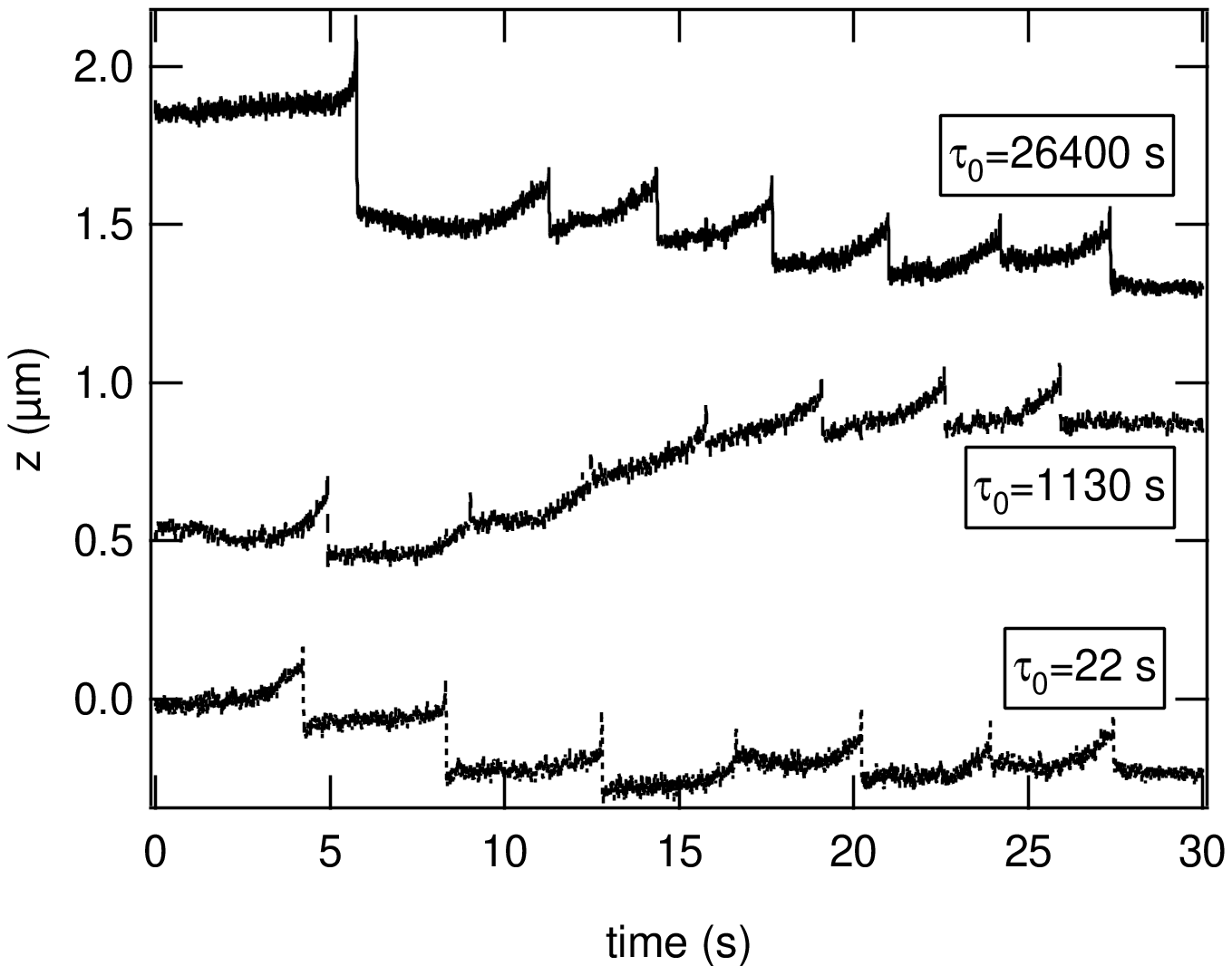,width=\linewidth}
\end{center}
\caption{Dilation during stick-slip motion (dry).
Vertical plate displacement vs time at $v=28.7 {\rm \mu m/s}$,
 $k=189.5 {\rm N/m}$. The motor is stopped during stick-slip motion
and restarted at $t=1.19 {\rm s}$ after a waiting time of $22$~s 
(dotted line), $1130$~s (dashed line), or $26400$~s (solid line).
The compaction following the first slip increases with waiting time;
this indicates that dilation occurred  during the waiting time.}
\end{figure}

\begin{figure} \begin{center}
\epsfig{file=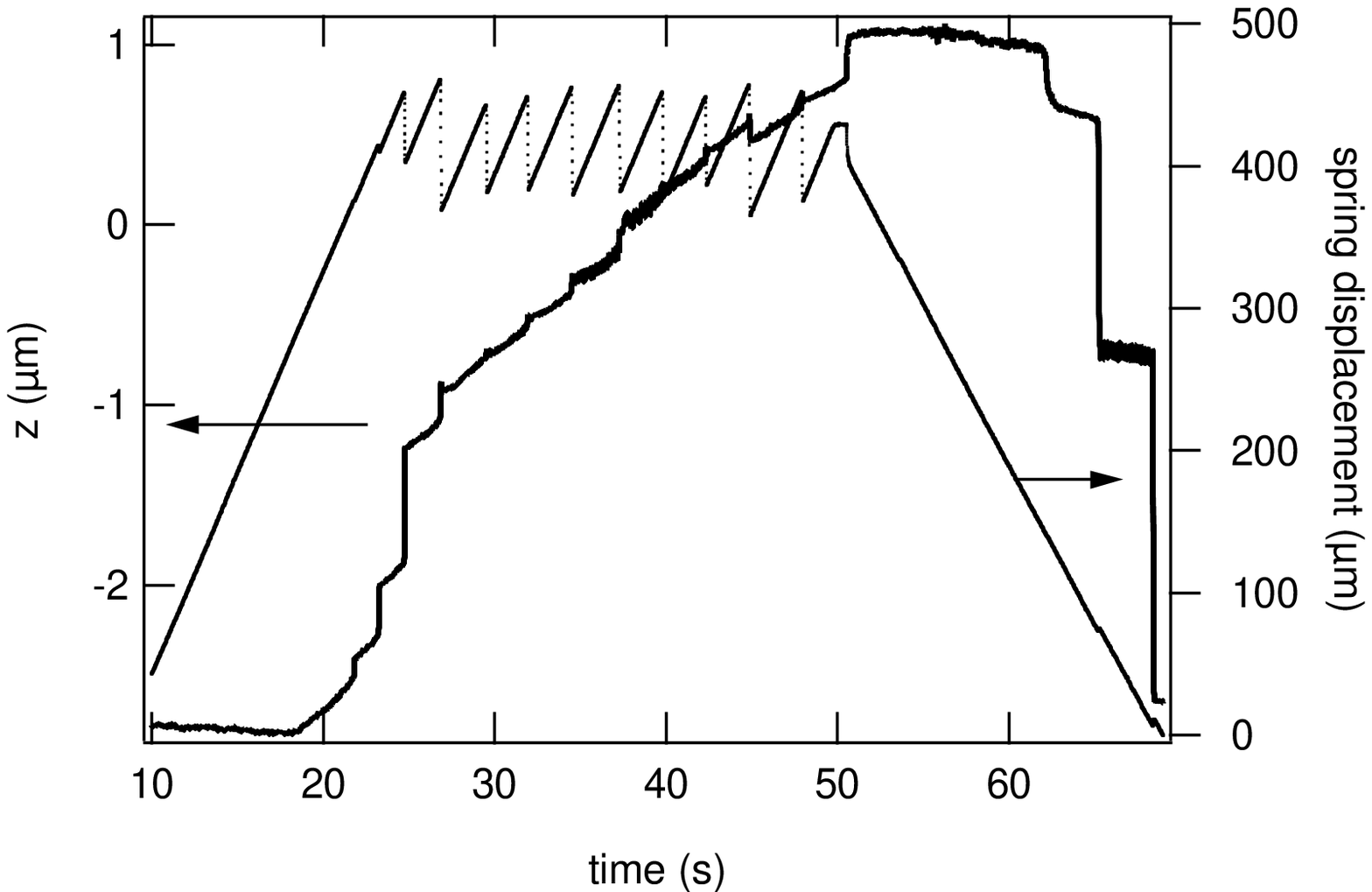,width=\linewidth}
\end{center}
\caption{Vertical displacement vs time (solid line)  
and spring displacement (dotted line) for the data of Figure~4
($v=28.7 {\rm \mu m/s}$, $k=189.5 {\rm N/m}$). The motor 
started with a completely unbent spring at $t=8.2$~s after 
a waiting time of $37213$~s. The material dilates when motion
starts.  
The reduction of the applied stress is accompanied by compaction.}
\end{figure}

\begin{figure} \begin{center}
\epsfig{file=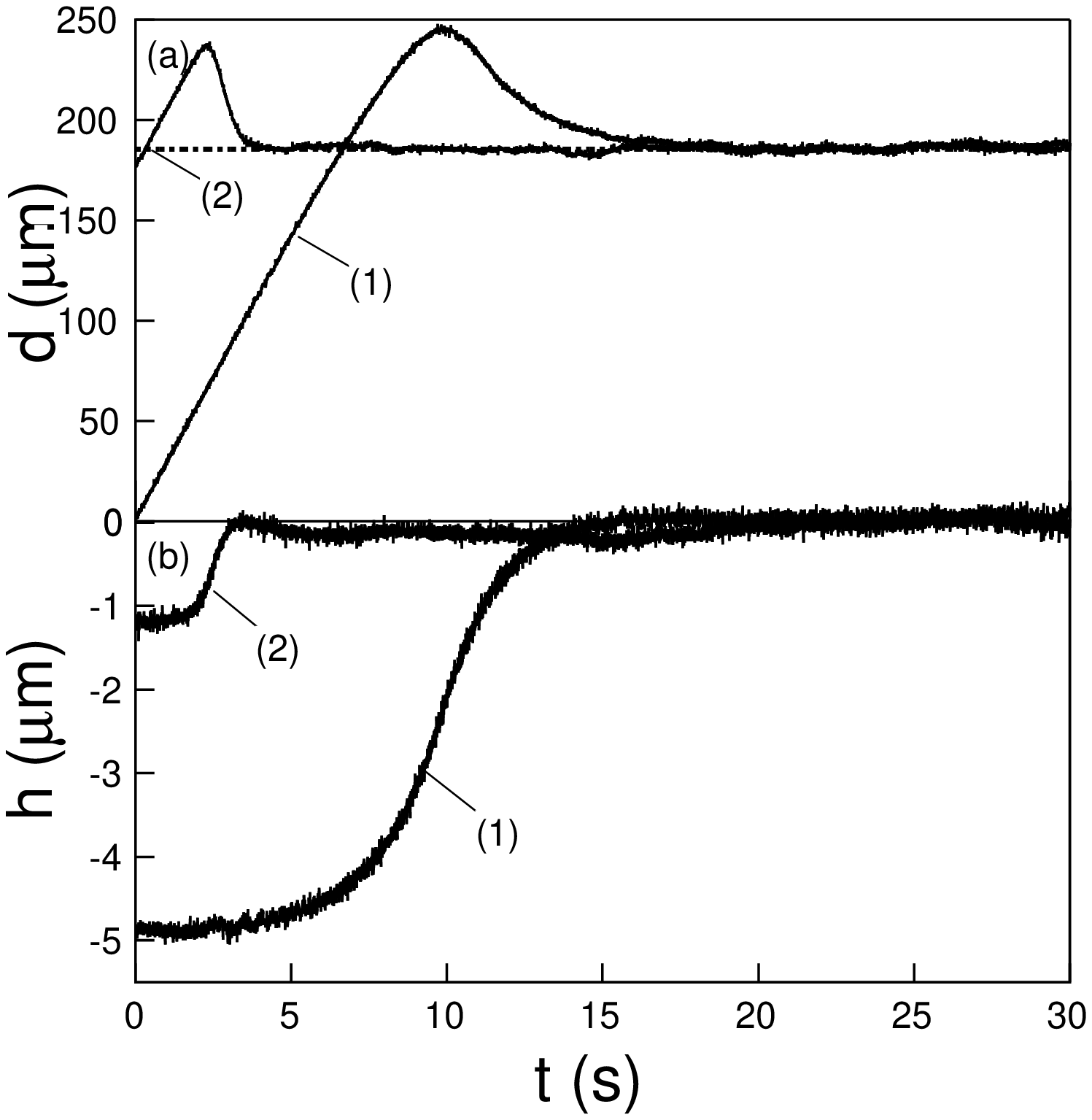,width=\linewidth}
\end{center}
\caption{(Wet friction) Behavior (a) of the spring displacement $d(t)$ and
(b) of the vertical position $h(t)$ as functions
of time $t$ in two different cases: (1) The horizontal stress
is released before the experiment;
(2) The horizontal stress is continuously applied.
In the second case, the layer is initially less packed because of the
applied shear stress; 
as a consequence, the total dilation $\Delta h$ observed during 
the experiment is less
($k = 189.5~\rm N/m$, $M = 14.5~{\rm g}$, $V = 28.17~{\rm \mu m/s}$)
(from Ref.[6]). }
\end{figure}

\begin{figure} 
\begin{center} 
\epsfig{file=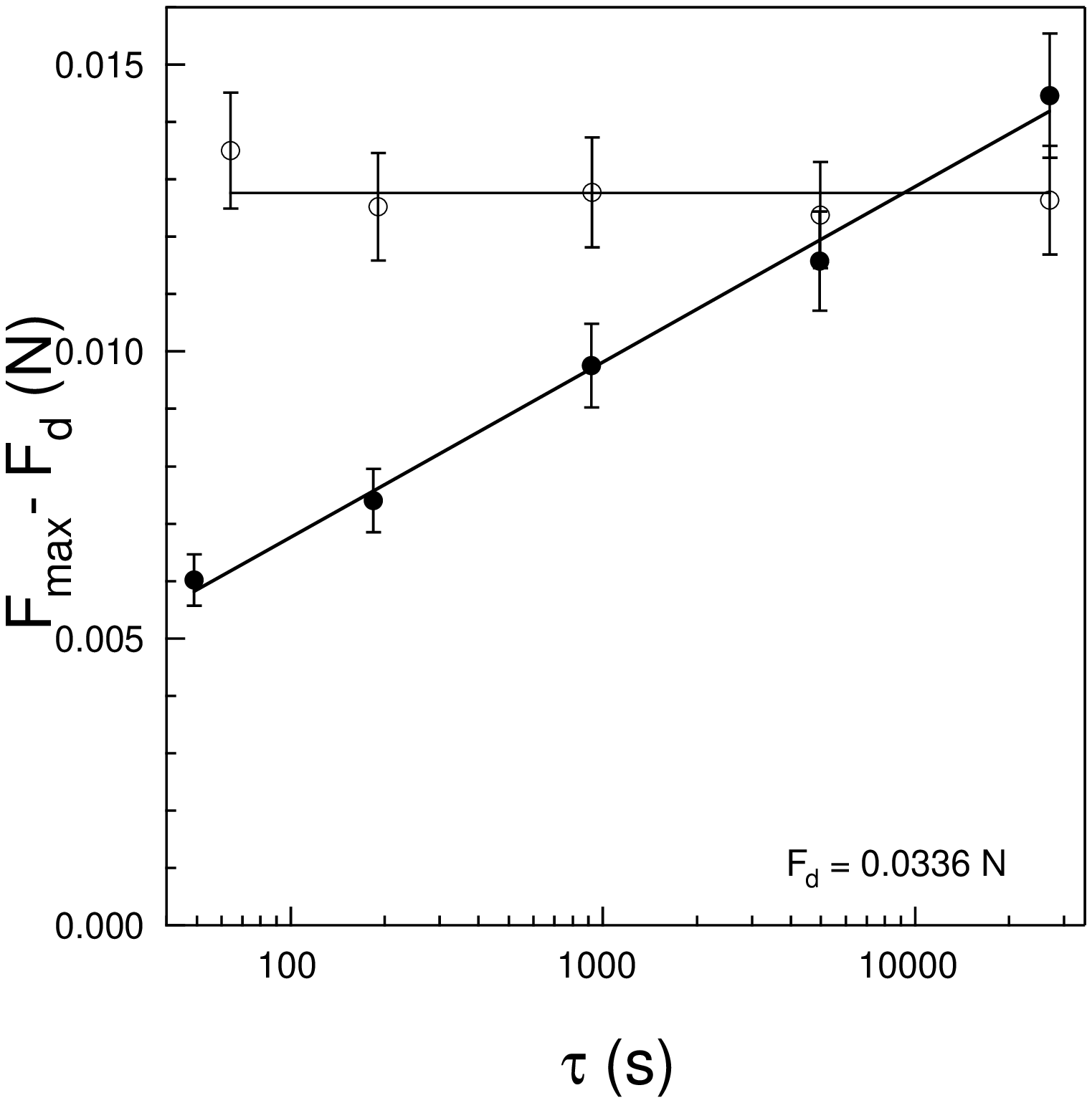,width=\linewidth}
\end{center}
\caption{(Wet friction) Maximum frictional force $F_{max}$ as a function 
of the waiting time $\tau$
($k = 189.5 {\rm N/m}$, $M = 14.5 {\rm g}$, $V = 28.17 {\rm \mu m/s}$).
Empty circles: no horizontal stress
is applied during the waiting time $\tau$. Filled circles: the plate
is submitted to a horizontal stress ($F \simeq F_d$)
during the waiting-time $\tau$.}
\end{figure}

\begin{figure} \begin{center}
\epsfig{file=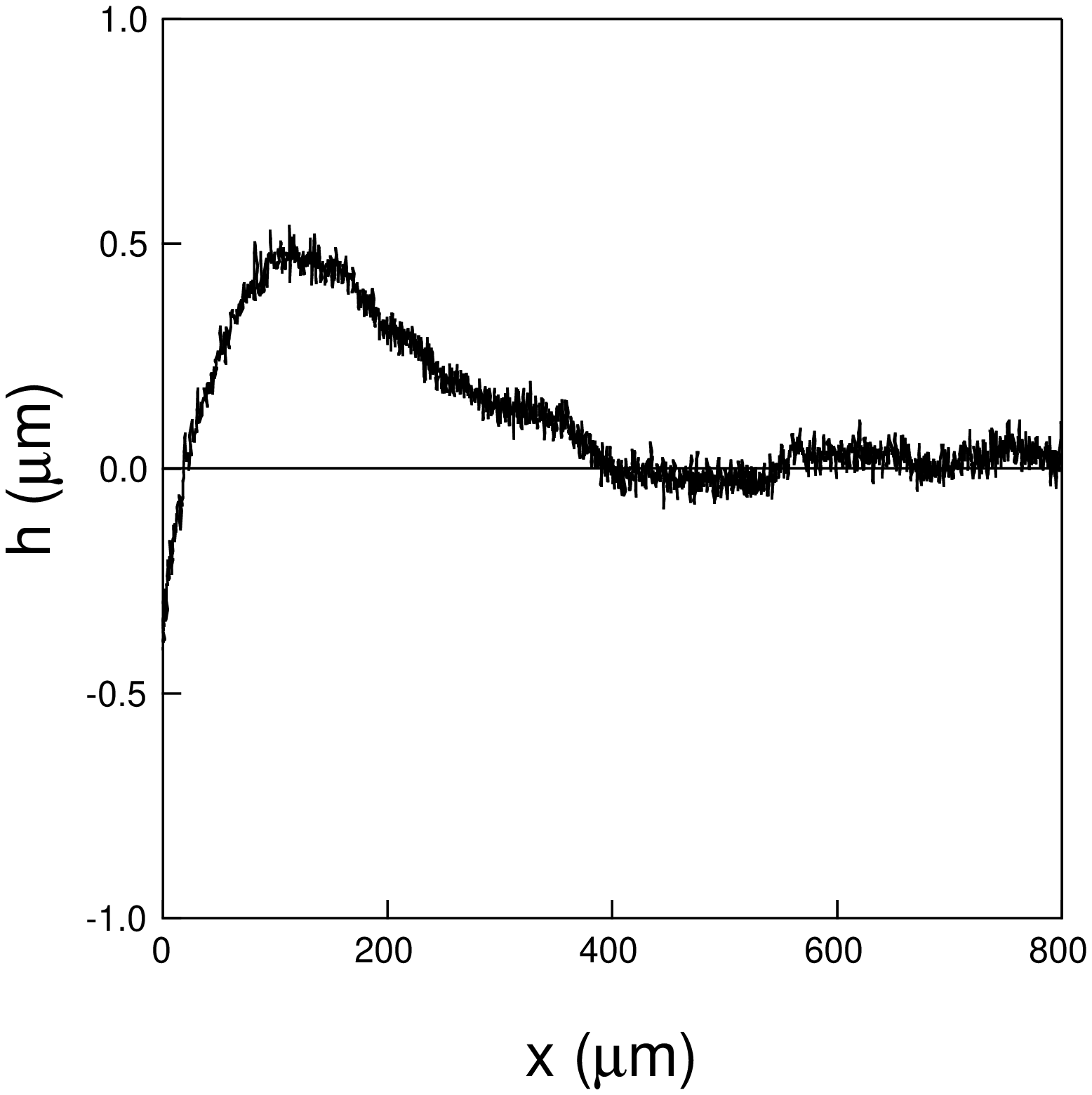,width=\linewidth}
\end{center}
\caption{(Wet friction) 
Vertical position of the plate $h$ as a function of its
horizontal position $x$ during a transient.
The vertical position overshoots when 
the plate was subjected to a horizontal
stress during the waiting time $\tau$
($k = 189.5 {\rm N/m}$, $M = 14.5 {\rm g}$, $V = 28.17 {\rm \mu m/s}$,
$\tau = 26000 {\rm s}$).}
\end{figure}

\begin{figure}\begin{center}
\epsfig{file=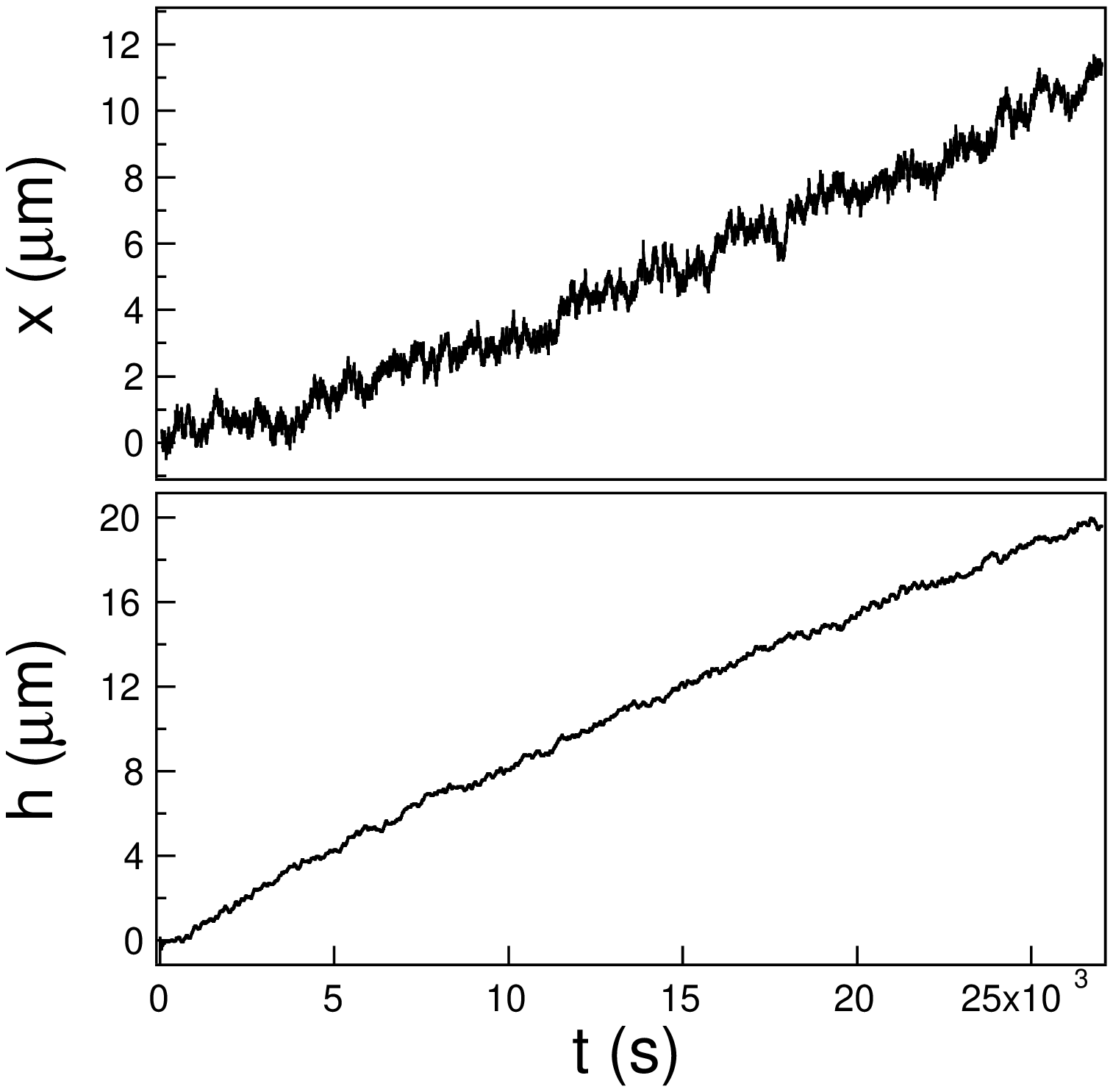,width=\linewidth}
\end{center}
\caption{(Wet friction)
Horizontal and vertical positions of the plate $x$ 
and $h$ as functions of time $t$ when the system is submitted
to a static horizontal applied stress $F \simeq F_d$. Creep of
about $1{\rm \mu m/h}$ is clearly evident
($k = 189.5 {\rm N/m}$, $M = 14.5 {\rm g}$).}
\end{figure}


\begin{thebibliography}{99}
\bibitem{footnote1} The contact area is calculated in the approximation 
that all particles in a layer sustain a roughly equal load.
\bibitem{footnote2} At the low speeds of our experiments, 
the plate does not move much after the motor has been stopped. The shear force
during waiting is therefore approximately equal $F_d$.
\end{thebibliography}
\end{document}